\documentclass[amssymb,twocolumn]{aastex63}

\usepackage{amsmath}
\usepackage{textcomp}
\usepackage{gensymb}%using \degree
\usepackage[hang,flushmargin,stable]{footmisc} 
\usepackage{graphicx}
\usepackage{txfonts}
\usepackage{xcolor}
\usepackage{float}
\usepackage{hyperref}
\usepackage{bm}
\usepackage{tabularx}
\usepackage{multirow}
\usepackage{lineno}

\newcommand{\orcid}[1]{\href{https://orcid.org/#1}{\includegraphics[width=8pt]{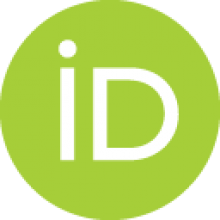}}}

\shorttitle{CV Evolution with Different MB}
\shortauthors{Tang et al. }
\graphicspath{{./}{Figures}}
\begin{document}

%\linenumbers
\title{The Evolution of Cataclysmic Variables Under Various Magnetic Braking Prescriptions}

\author{Wen-Shi Tang\orcid{0000-0002-6588-9264}}
%\affiliation{School of Astronomy and Space Science, Nanjing University, Nanjing 210023, China; \href{mailto:tangwenshi20@163.com}{tangwenshi20@163.com}}%\href{mailto:lixd@nju.edu.cn}{lixd@nju.edu.cn}}
%\affiliation{Key Laboratory of Modern Astronomy and Astrophysics (Nanjing University), Ministry of Education, Nanjing 210023, China}
\affiliation{Department of Astronomy, Xiamen University, Xiamen, Fujian 361005,China; \href{mailto:tangwenshi20@163.com}{tangwenshi20@163.com}}

\author{Xiang-Dong Li\orcid{0000-0002-0584-8145}}
\affiliation{School of Astronomy and Space Science, Nanjing University, Nanjing 210023, China; \href{mailto:lixd@nju.edu.cn}{lixd@nju.edu.cn}}
\affiliation{Key Laboratory of Modern Astronomy and Astrophysics (Nanjing University), Ministry of Education, Nanjing 210023, China}

%\author{Bo Wang\orcid{https://orcid.org/0000-0002-3231-1167}}
%\affiliation{Yunnan Observatories, Chinese Academy of Sciences, Kunming 650216, China}
%\affiliation{International Centre of Supernovae, Yunnan Key Laboratory, Kunming, 650216, China}

\author{Zhe Cui\orcid{0000-0001-8311-0608}}
%\affiliation{School of Astronomy and Space Science, Nanjing University, Nanjing 210023, China; \href{mailto:tangwenshi20@163.com}{tangwenshi20@163.com},\href{mailto:lixd@nju.edu.cn}{lixd@nju.edu.cn}}
%\affiliation{Key Laboratory of Modern Astronomy and Astrophysics (Nanjing University), Ministry of Education, Nanjing 210023, China}
\affiliation{College of Physics and Electronic Information, Dezhou University, Dezhou 253023, China}
\affiliation{Shandong Key Laboratory of Space Environment and Exploration Technology, College of Physics and Electronic Information, Dezhou University, Dezhou 253023, China}

\author{Zhu-ling Deng\orcid{0000-0001-8311-0608}}
%\affiliation{School of Astronomy and Space Science, Nanjing University, Nanjing 210023, China; \href{mailto:tangwenshi20@163.com}{tangwenshi20@163.com}}%\href{mailto:lixd@nju.edu.cn}{lixd@nju.edu.cn}}
\affiliation{Yunnan Observatories, Chinese Academy of Sciences, Kunming 650216, China}
%\affiliation{College of Physics and Electronic Information, Dezhou University, Dezhou 253023, China}
\author{Wei-Min Gu}
%\affiliation{School of Astronomy and Space Science, Nanjing University, Nanjing 210023, China; \href{mailto:tangwenshi20@163.com}{tangwenshi20@163.com}}%\href{mailto:lixd@nju.edu.cn}{lixd@nju.edu.cn}}
%\affiliation{Key Laboratory of Modern Astronomy and Astrophysics (Nanjing University), Ministry of Education, Nanjing 210023, China}
\affiliation{Department of Astronomy, Xiamen University, Xiamen, Fujian 361005,China; \href{mailto:tangwenshi20@163.com}{tangwenshi20@163.com}}
%\author{Belloni\orcid{0000-0001-8311-0608}}
%\affiliation{School of Astronomy and Space Science, Nanjing University, Nanjing 210023, China; \href{mailto:tangwenshi20@163.com}{tangwenshi20@163.com}}%\href{mailto:lixd@nju.edu.cn}{lixd@nju.edu.cn}}
%\affiliation{Key Laboratory of Modern Astronomy and Astrophysics (Nanjing University), Ministry of Education, Nanjing 210023, China}
%\affiliation{College of Physics and Electronic Information, Dezhou University, Dezhou 253023, China}

\begin{abstract}

Recent studies revealed discrepancies between observations and the predictions of the standard magnetic braking (MB). Although alternative models have been broadly discussed in neutron star binaries, they have not been systematically tested in cataclysmic variables (CVs). In this work, we investigate the performance of four MB models in CVs: the standard MB, the Convection And Rotation Boosted (CARB) model, the $\tau$-boosted model, and the saturated, boosted, and disrupted (SBD) model. We find that both the CARB and $\tau$-boosted models appear too strong so that it fails to reproduce the location of the period gap in CVs, indicating that they are not appropriate for CVs. Furthermore, we present a comparison between the standard MB and the SBD models. Compared with the standard model, although the SBD model can better reproduce some observational features, it also exacerbates certain discrepancies between theory and observations. We also find that different prescriptions for the convective turnover timescale have a significant impact on the results in the non-standard MBs. Finally, we discuss the impact of the SBD model on the formation and evolution of AM CVn.

\end{abstract}


\section{Introduction}\label{Sect_Intro}
Angular momentum loss plays a key role in the formation of a wide range of astronomical phenomena, including cataclysmic variables (CVs), X-ray binaries, and so on. The angular momentum loss mechanisms mainly include gravitational wave radiation, mass loss and magnetic braking (MB).  For binaries with unevolved main sequence (MS) companions, e.g. CVs, MB is the dominant mechanism driving orbital shrinkage. MB originates from the coupling between the magnetic field and stellar winds of the companion. \citep{1968MNRAS.138..359M, 1981A&A...100L...7V}. In a binary system, if tidal forces synchronize the spin of the companion star with the orbit, magnetized stellar winds, which remove stellar spin angular momentum, effectively extract orbital angular momentum from the binary. 

A widely used MB formalism is contributed from \cite{1983ApJ...275..713R}. It has become established as the standard model in the field of binary evolution simulations. A key feature of this formalism is its ability to produce the period gap in CVs. It has been extended to simulate the evolution of neutron star/black hole low-mass X-ray binaries (NS/BH LMXBs). Unfortunately, increasing evidence suggests that the predictions of the standard model do not fully agree with the observations. The observed mass accretion rates of NS LMXBs are about 1-2 orders of magnitude higher than that predicted by the standard model \citep{2002ApJ...565.1107P, 2003ApJ...597.1036P, 2015ApJ...809...99S, 2019MNRAS.483.5595V, 2021ApJ...909..174D}. The predicted orbital period distribution of binary millisecond pulsars by the standard model is not aligned with the observation \citep{2003ApJ...597.1036P, 2014A&A...571A..45I, 2015ApJ...809...99S}. Moreover, the standard model also faces a fine-tuning problem in the formation of ultra-compact binaries, e.g. ultra-compact X-ray binary and binary millisecond pulsars with extremely low-mass white dwarfs (ELM WDs; $M_{\rm WD}\lesssim 0.2\,M_{\rm \odot}$), since only a very narrow range of initial orbital periods can form such systems \citep[e.g. ][]{2014A&A...571A..45I}. Regarding CVs, the standard MB can not perfectly reproduce the mass$-$radius relation of the main sequence companions and the period gap in CVs, unless a weaker MB  (e.g., a scaling factor of 0.66 for the standard MB) is adopted \citep{2011ApJS..194...28K, 2025ApJ...990..141T}.

 Motivated by the above discrepancies, several modified MB models have been proposed in recent years \citep{2012ApJ...754L..26M,2012ApJ...746...43R, 2019MNRAS.483.5595V, 2019ApJ...886L..31V}. Among these models, those most widely discussed in recent years in the context of binary evolution include the $\tau$-boosted model and Convection And Rotation Boosted (CARB) model. The $\tau$-boosted model was initially developed by \cite{2019MNRAS.483.5595V}. This prescription includes scaling of the magnetic field strength with the convective turnover time of the companion. \cite{2021ApJ...909..174D} showed that, when compared with NS LMXBs and binary pulsars, the $\tau$-boosted model can provide the best match to the observed properties. This model also contributes to alleviating some of the existing problems, such as the fine-tuning problem for the formation of ultra-compact binary and the formation of the transient NS LMXB Swift J1858.6‑0814 \citep{2021MNRAS.503.3540C, 2024MNRAS.530.4277E, 2024ApJ...974..298Y}. 
 
The CARB model was put forward by \cite{2019ApJ...886L..31V}. It is self-consistently derived by accounting for the effects of stellar rotation on the Alfven radius and the magnetic field dependence on convective turnover time. \cite{2019ApJ...886L..31V} pointed out that the CARB model can reproduce some key observational properties of all persistent LMXBs, including  mass transfer rate, orbital period, mass ratio and effective temperature. A number of subsequent studies have shown that the CARB model is capable of resolving several issues, such as  the fine-tuning problem in the formation of ultra-compact binaries \citep{2021MNRAS.506.3266S,2023A&A...678A..34B}, the orbital evolution of some peculiar LMXBs \citep{2023A&A...679A..74W, 2024ApJ...976..210F}, the formation of extremely long-period CVs \citep{2025arXiv250821358T} and SDSS J1257+5428 \citep{2025A&A...697A.100B}.

Although these prescriptions have been successful in neutron star systems, their validity in other systems remains uncertain. For example, \cite{2024ApJ...971...54D} recently pointed out that, contrary to what was found for NS LMXBs requiring strong MB, the MB laws with relatively low efficiency (e.g. standard MB or the form suggested by \cite{2012ApJ...746...43R}) appears to be more consistent with the characteristics of black hole binaries. \cite{2024ApJ...974..298Y} also pointed out that the efficiency of MB may vary among different systems, with certain prescriptions being more suitable for specific types. Furthermore, these prescriptions have so far not been examined in CVs. Apart from MB, other forms of angular-momentum loss do not seem to naturally account for the period gap in CVs \citep{1995ApJ...439..330K}. Therefore, it is essential to test the MB prescriptions in CVs, particularly to examine whether they can reproduce the observed period gap. This is one of the goals of this paper. 

It is noteworthy that a model referred to as saturated, boosted, and disrupted (SBD) MB has recently been proposed (\citealt{2024A&A...682A..33B,2025A&A...696A..92B}). The idea for the SBD model is that recent studies have offered strong evidence that magnetic activity reaches a saturation level when the stellar rotation is faster than a critical value (\citealt{2020ApJ...905..107M}). If this relates with MB, it implies a much shallower dependence of the torque on the spin period below a given rotation period \citep{2009ApJ...692..538R,2020A&A...638A..20M,2020ApJ...905..107M}. In addition, \cite{2022MNRAS.517.4916E} found that the observed orbital period distributions of  low-mass detached main-sequence eclipsing binaries are best reproduced by MB models in which the magnetic field saturates at short period. Based on these, \cite{2024A&A...682A..33B} proposed a MB model that includes magnetic saturation. The model predicts sufficiently strong angular-momentum loss for stars with radiative cores, a drastic reduction at the fully convective boundary. \cite{2025A&A...696A..92B} applied the SBD model to the evolution of CVs and calibrated it using the period gap. However, they applied the SBD model only to a few specific evolutionary tracks. In this paper, we utilize the SBD model for a binary population synthesis study.

 This paper is organized as follows. In Section \ref{Sect_method}, we describe our methods. In Section \ref{Sect-results}, we present our results and provide some discussion in Section \ref{Discuss}. Lastly, summaries are given in Section \ref{Sect-Sum}.
 
\section{Assumptions and Methods}\label{Sect_method}
\subsection{Basic assumption for detailed binary evolution calculations}\label{subsection-Basic assumption}
We use the stellar evolution code {\tt MESA} \citep[version 11701; ][]{2011ApJS..192....3P, 2013ApJS..208....4P,2015ApJS..220...15P,2019ApJS..243...10P} to evolve a large number of WD+MS binaries. The initial mass of WD  ($M_{\rm WD,i}$) is set to be $0.4-1.2\,M_{\rm \odot}$ in a step of $0.2\,M_{\rm \odot}$. The initial companion masses ($M_{\rm2, i}$) range from $0.4\,M_{\rm \odot}$ to $1.5\,M_{\rm \odot}$ in a step of $0.2\,M_{\rm \odot}$. The logarithmic orbital period ($\log_{10}(P_{\rm orb, i}$/days)) is -0.5 to 2.0 in a step of 0.03. The angular momentum loss includes gravitational wave radiation \citep{1975ctf..book.....L}, mass loss and MB. The treatment of angular-momentum loss due to mass loss is described below, while MB is discussed in Section \ref{subsection-MBs}.

Depending on the mass ratio, some systems could experience a thermal-timescale mass transfer phase at the beginning of mass transfer, potentially resulting in mass accumulation onto accreting WDs. In our calculation, we include this possible mass accretion by WDs.  The mass growth rate of the WD is expressed as

\begin{equation}
\dot M_{\rm WD} =  \eta_{\rm H}\eta_{\rm He} |\dot M_{\rm 2}|,
\end{equation}
where $\eta_{\rm H}$ and $\eta_{\rm He}$ are the mass accumulation efficiencies for hydrogen and helium burning, respectively. $\dot M_{2}$ is mass transfer rate of the companion. $\eta_{\rm He}$ is adopted from the result of \cite{2017A&A...604A..31W}. For $\eta_{\rm H}$, we take the same form  in  \cite{2010MNRAS.401.2729W},
 \begin{equation}
\eta _{\rm H}=\left\{
 \begin{array}{ll}
 \dot{M}_{\rm crit}/|\dot{M}_{\rm 2}|, & {\rm if}\ |\dot{M}_{\rm 2}|> \dot{M}_{\rm crit},\\
 1, & {\rm if}\ \dot{M}_{\rm crit}\geq |\dot{M}_{\rm 2}|\geq \dot{M}_{\rm stable},\\
 0, & |\dot{M}_{\rm 2}|< \dot{M}_{\rm stable}.
\end{array}\right.
\end{equation}
$\dot M_{\rm stable}$ and $\dot M_{\rm crit}$ are taken from \cite{2013ApJ...778L..32M} ( see also \citealt{2024ApJ...977...34T}). The material that cannot be retained by the WD leaves the binary in the form of optically thick winds, carrying away the specific orbital angular momentum of the WD \citep{2001ApJ...563..958K}. We do not include the empirical consequential angular momentum loss (eCAML) proposed by \cite{2016MNRAS.455L..16S}, but see \cite{2025A&A...696A..92B} for a discussion about the effect of eCAML on the predicted properties of CVs.

\subsection{Magnetic braking}\label{subsection-MBs}
We adopt four MB models, including the standard MB, $\tau$-boosted, CARB and SBD models. Their detailed expressions are listed as below.

\subsubsection{Standard model}
\cite{1972ApJ...171..565S} found that the rotational periods of Sun-like stars are proportional to the square root of their age, demonstrating that they spin down as they evolve. This spin-down arises from the coupling of the stellar wind to magnetic fields. Later,  \cite{1983ApJ...275..713R} extended this law to binary systems and presented a formula for angular momentum loss rate of MB (hereafter RVJ; see also \citealt{2015ApJS..220...15P}),  

\begin{equation}
\dot J_{\rm MB, RVJ} =  -6.82\times 10^{34} \left( \frac{M_{2}}{M_{\rm \odot}}\right) \left( \frac{R_{2}}{R_{\rm \odot}}\right)^{\gamma_{\rm MB}}\left( \frac{1\rm d}{P_{\rm orb}}\right)^{3} [\rm dyn\,cm],
\end{equation}
 where $R_2$ is the radius of the companion, the subscript `$_\odot$' denotes solar values. $\gamma_{\rm MB}$ is a dimensionless parameter and is set to be 3 (the default value in {\tt MESA}). 
 
 \subsubsection{$\tau-$boosted model}
 \cite{2019MNRAS.483.5595V} derived a modified form of MB, whose final expression can be regarded as the standard law including a scaling of the magnetic field strength with the convective turnover time $\tau_{\rm conv}$, wind mass-loss rate $\dot M_{\rm 2, wind}$ and rotation rate $\Omega$ of the companion,
 \begin{equation}\label{Eq-tau-boosted}
\dot J_{\rm MB, \tau-boosted} = \dot J_{\rm MB,RVJ}\left( \frac{\Omega}{\Omega_{\rm \odot}}\right)^{\beta}  \left( \frac{\tau_{\rm conv}}{\tau_{\rm \odot, conv}}\right)^{\xi} \left( \frac{\dot M_{\rm 2, wind}}{\dot M_{\rm \odot, wind}}\right)^{\alpha},      
\end{equation}
where $\Omega_{\rm \odot}\sim 3\times 10^{-6}\rm s^{-1}$, $\tau_{\rm \odot, conv}=2.8\times 10^6\,\rm s$. The wind mass-loss rate is calculated based on \cite{1975MSRSL...8..369R}.

 From Eq.\,(\ref{Eq-tau-boosted}), the precise formulation of MB is determined by the exponent ($\beta, \xi, \alpha$). \cite{2019MNRAS.483.5595V} defined three additional MB cases in addition to the standard one (corresponding to ($\beta, \xi, \alpha$) = (0, 0, 0)). These three cases are named respectively as: `Convective ($\tau$) $-$ boosted' with ($\beta, \xi, \alpha$) = (0, 2, 0), ‘Intermediate’ with ($\beta, \xi, \alpha$) = (0, 2, 1), `Wind-boosted' with ($\beta, \xi, \alpha$) = (2, 4, 1). Since \cite{2019MNRAS.483.5595V} and \cite{2021ApJ...909..174D} showed that the $\tau$-boosted model more effectively reproduces properties of NS LMXBs and binary pulsars , we restrict our discussion to this model.
 
 \subsubsection{CARB model}
 Considering the dependence of the magnetic field strength on the outer convective zone and the dependence of the Alfven radius on donor’s rotation, \cite{2019ApJ...886L..31V} proposed a modified MB model, referred to as Convection And Rotation Boosted (CARB), 

 \begin{eqnarray}\label{Eq-CARB}
 \dot{J}_{\mathrm{MB, CARB}} 
= & -\frac{2}{3} \dot{M}_{\mathrm{2, wind}}^{-1/3} R_{2}^{14/3}
\left( v_{\mathrm{esc}}^{2} + \frac{2 \Omega^{2} R_{2}^{2}}{K_{2}^{2}} \right)^{-2/3} \nonumber \\
& \times \, \Omega_{\odot} B_{\odot}^{8/3}
\left( \frac{\Omega}{\Omega_{\odot}} \right)^{11/3}
\left( \frac{\tau_{\mathrm{conv}}}{\tau_{\odot,\mathrm{conv}}} \right)^{8/3},
\end{eqnarray}
where $v_{\rm esc}$ is the surface escape velocity of the companion. $B_{\rm \odot}$ ( = 1\,G) is the magnetic field strength on the surface of the Sun and $K_{2}$=0.07. Other parameters have the same meanings as described above.

\subsubsection{Saturated, boosted, and disrupted model}
In rapidly rotating low-mass stars, magnetic activity (such as X-ray and chromospheric activity) saturates once a critical angular velocity is reached. This causes the relationship between the MB torque and the rotational velocity to flatten. As a result, MB may become less dependent on the rotation rate at short period, a phenomenon known as `saturation'. The saturated MB can be expressed as (\citealt{1995ApJ...441..865C})
\begin{equation}\label{Eq-Jdot-SAT}
\dot J_{\mathrm{SAT}} = -\beta \left( \frac{R_2}{\mathrm{R}_{\odot}} \frac{\mathrm{M}_{\odot}}{M_2} \right)^{1/2} 
\begin{cases} 
\Omega^3, & \text{if } \Omega \leq \Omega_{\mathrm{crit}}, \\
\Omega\,\Omega_{\mathrm{crit}}^2, & \text{if } \Omega > \Omega_{\mathrm{crit}},
\end{cases}
\end{equation}
where $\beta=2.7\times 10^{47} \rm erg\,s^{-1}$ \citep{2003ApJ...582..358A}. $\Omega_{\rm crit}$ is the critical angular velocity above which the rotation of star achieves saturation.  $\Omega_{\rm crit}$ is assumed to be 
\begin{equation}
\Omega_{\rm crit} = 10\,\Omega_{\rm \odot} \left( \frac{\tau_{\rm \odot}}{\tau_{\rm conv}}\right).
\end{equation}

The saturation MB has been confirmed in low-mass detached eclipsing binaries \citep{2022MNRAS.517.4916E} and detached close binaries with M and K dwarfs \citep{2024A&A...682A..33B}.
%Based on this, \cite{2024A&A...682A..33B} suggested a magnetic braking prescription that predicts strong enough angular momentum loss for stars that still have a radiative core, a drastic decrease at the fully convective boundary, and a period dependence that includes saturation.
To explore the effect of different strengths of MB and different degrees of disruption at the fully convective boundary, two parameters $K$ and $\eta$ are introduced by \cite{2024A&A...682A..33B},
\begin{equation} \label{Eq-SBD}
\dot J_{\mathrm{MB, SBD}} = 
\begin{cases} 
K \, \dot J_{\mathrm{SAT}}, & \text{(radiative core + convective envelope),} \\
\left( K \, \dot J_{\mathrm{SAT}} \right) / \eta, & \text{(fully convective).}
\end{cases}
\end{equation}

Eq.\,(\ref{Eq-SBD}) is referred to as the saturated, boosted, and disrupted (SBD) MB. \cite{2025A&A...696A..92B} applied Eq. \,(\ref{Eq-SBD}) to CV evolution. In their simulations, both the values of $K$ and $\eta$ are set to be constant and vary from 20 to 80. They found that to reproduce the period gap and minimum orbital period in CVs, the best fitting result is $K\sim 30$ and $\eta\sim 20$.

From Equations~(\ref{Eq-tau-boosted}) to (\ref{Eq-SBD}), it is clear that the $\tau$-boosted, CARB, and SBD models all depend on $\tau_{\rm conv}$. At present, various forms of the convective turnover timescale exist in the literature, and different studies adopt different calculation methods \citep{2019MNRAS.483.5595V, 2021ApJ...909..174D, 2025A&A...696A..92B}. As we will demonstrate in Section\,\ref{Effect-tau}, different prescriptions for $\tau_{\rm conv}$ can have a significant impact on the results. In the main text of our study, however, we adopt the prescription of \cite{2025ApJ...988..102G}, as it provides good agreement with observational constraints. Specifically, \cite{2025ApJ...988..102G} calculated a theoretical local $\tau_{\rm conv}$ at a position $r$ as %(Their code is available on \url{https://zenodo.org/records/15680676})}
\begin{equation}\label{Eq-tau-Gossa}
\tau_{\rm conv}(r)= \frac{H_{\rm p}(r)}{v_{\rm }(r)},
\end{equation}
where $v(r)$ is the local convective velocity, and $H_{\rm P}(r)$ is the local pressure scale height. \cite{2025ApJ...988..102G} evaluate both $v(r)$ and $H_{\rm P}(r)$ at a location defined as the base of the convective envelope (BCE) plus half of a local pressure scale height, i.e., $r_{H_{\rm P}}(r) = r_{\rm BCE} + 0.5\,H_{\rm P}(r)$. Rather than simply setting $r = r_{\rm BCE}$, they solve $r_{\rm i}\lesssim r_{\rm BCE}+0.5H_{\rm P}(r_{\rm i})$ for each $i$\,th model cell of stellar structure. Iteration is stopped once the condition is satisfied, and then let $r=r_i$. \cite{2025ApJ...988..102G} showed that by adopting such a method, they can reproduce the observed $\tau_{\rm conv}$ of low-mass stars, in particular the bump for stars with mass in the range of $0.35-0.4\,M_{\rm \odot}$.

\section{Results}\label{Sect-results}
\begin{figure*}[ht]
\centering
\includegraphics[width=18cm,height=7cm]{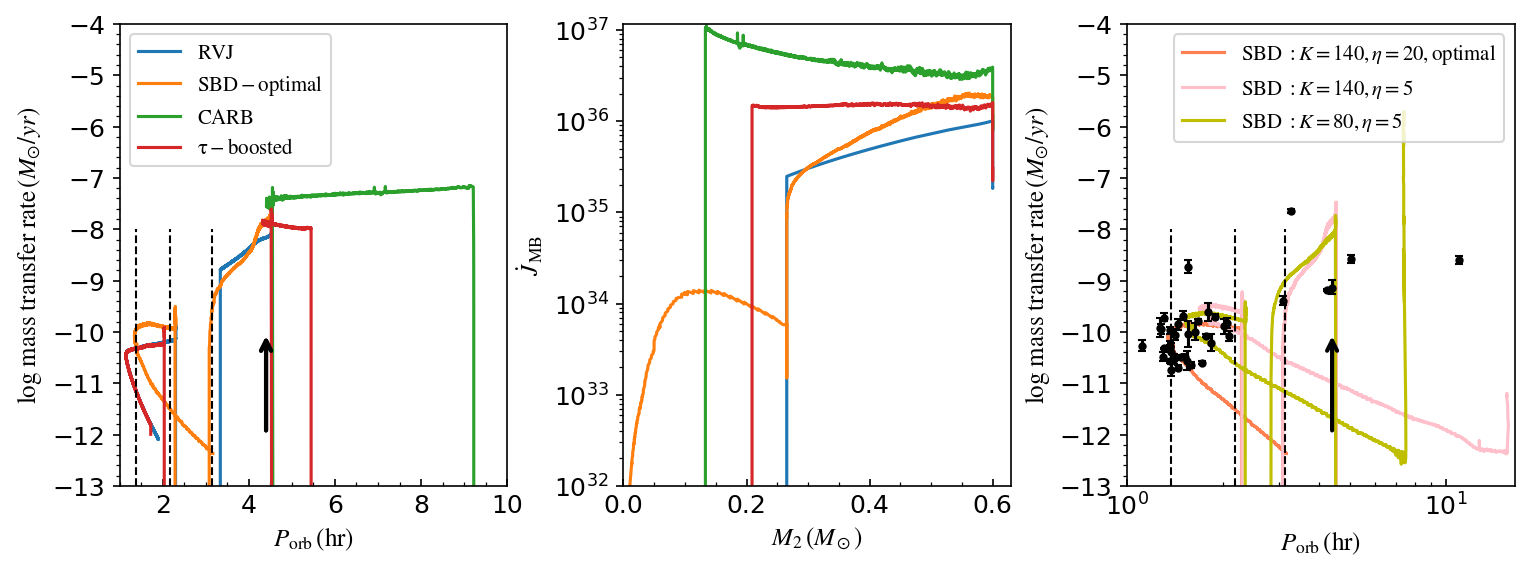}
 \caption{An example of effect of different MBs on binary evolution. The initial systems have parameters ($M_{\rm WD,i}/M_{\rm \odot}$, $M_{\rm 2,i}/M_{\rm \odot}$, $P_{\rm orb,i}/\rm days$) = (0.8, 0.6, 0.339), and they eventually evolve into a CV.  The left panel presents the evolution of the mass transfer rate as a function of orbital period, while middle panel shows the angular momentum loss rate of MB as a function of companion mass. In each panel, the blue, yellow, green and red lines correspond to the results of RVJ, SBD, CARB and $\tau-$boosted models, respectively. For the SBD model, we set $K=140$ and $\eta = 20$. The three vertical dashed lines stand for the boundaries of the period gap (2.15-3.18\,hr) and the minimum orbital period (82.4\,min) \citep{2011ApJS..194...28K}. The right panel shows the effect of $K$ and $\eta$ on the binary evolution for the SBD model. The black dots are the observational data from \cite{2022MNRAS.510.6110P}. Note that in the left panel, the evolution of the $\tau$-boosted model overlaps with that of the RVJ model below the period gap, while in the right panel the SBD model with $K=140$ and $\eta=20$ shows identical evolution to the model with $K=140$ and $\eta=5$ above the period gap. In the left and right panels, the vertical black arrow indicates the starting point and direction of the evolution track. In the middle panel, the evolution proceeds from right to left.}
\label{Fig-example1}
\end{figure*}

\subsection{Examples of the impact of different MBs on cataclysmic variable evolution} 
Figure\,\ref{Fig-example1} illustrates the influence of different MBs on the evolution of a binary system that eventually forms a CV. The left panel shows the evolution of the mass transfer rate as a function of orbital period. The mass transfer begins at about $P_{\rm orb}\sim 4.5\,\rm hr$ (the position of black arrow) for all models. The figure shows that after the onset of mass transfer, the evolution proceeds toward increasing orbital periods for both the CARB and $\tau-$boosted models. For the $\tau-$boosted model (the red line), the orbital period increases from $\sim 4.5$\,hr to $\sim 5.5\,\rm hr$. The MB stops as the companion becomes fully convective. Then the companion star fill its Roche lobe again under the gravitational wave radiation at $P_{\rm orb}\sim 2\,\rm hr$. Since then, the system's evolution is nearly identical to the RVJ model (the blue line). In contrast, under the CARB model, the orbital period increases from $\sim$4.5 hr to $\sim$9 hr during mass transfer, then MB ceases and system becomes detached. After that, even if the system evolves to a Hubble timescale, the companion star can no longer fill its Roche lobe. The expansion of orbit in these two models can be attributed to the high mass transfer rate induced by strong angular momentum loss  (see middle panel). As a result, both the CARB model and $\tau-$boosted model clearly fail to reproduce the location of the period gap in CVs. We conclude that these two models are unsuitable for CV evolution\footnote{For the CARB model, we have evolved a grid of binaries with initial parameters described in Section\,\ref{subsection-Basic assumption}, and found that all evolutionary tracks are similar to that shown in Figure\,1, thereby confirming this conclusion.  For the $\tau$-boosted model, the same conclusion can also be drawn from the analysis of \cite{2026A&A...707A..76Z} (see Section~\ref{Discus-comparison}).}, at least from the perspective of reproducing the period gap, and they will not be discussed further in the main text.

We now compare the RVJ model and the SBD model. To determine the optimal SBD prescription, we investigate the effects of different $K$ and $\eta$ on the CV evolution,  as illustrated in the right panel of Figure\,\ref{Fig-example1}. We find that a larger $K$ leads to a higher upper edge of the period gap, while within the explored range of $K$ and $\eta$, the lower boundary is only weakly affected. Meanwhile, a smaller $\eta$ results in higher mass transfer rates and a larger minimum orbital period value. For the cases with $\eta = 5$, the mass transfer rates at the end of the evolution show a sharp increase. This occurs because the envelope of the companion is nearly completely stripped\footnote{The final mass is smaller than $10^{-3}\,M_{\rm \odot}$, which is comparable to Jupiter.}, leading to a rapid decrease in $\tau_{\rm conv}$, since only the envelope-related quantity enters the calculation of $\tau_{\rm conv}$ in Eq.\,(\ref{Eq-tau-Gossa}). This, in turn, produces a sharp increase in the MB strength. In the figure, the observed mass transfer rates of CVs given by \cite{2022MNRAS.510.6110P} are also plotted. We find that a value of $K=140$ and $\eta =20$ can provide the best agreement with the orbital properties of CVs (the period gap and minimum period) and observed mass transfer rate below the period gap. Hereafter, we refer the SBD model with  $K=140$ and $\eta =20$ as the ‘SBD-optimal' model.

The evolutionary example under the SBD-optimal model is also drew in the left panel of Figure\,\ref{Fig-example1} for comparison with the RVJ model. It is seen that the RVJ model slightly overestimates the upper boundary of the period gap and underestimates the minimum orbital period, which is consistent with the findings of previous studies \citep[e.g. ][]{2011ApJS..194...28K}. In contrast, the SBD-optimal model almost perfectly reproduces the observed period gap and the minimum orbital period. However, it should be noted that the orbital period of period bouncers in the RVJ model expands to no more than $2\,\mathrm{hr}$, whereas in the SBD-optimal model the post-minimum-period evolution can cross the period gap again. This will lead to a large number of systems with sub-dwarfs residing in the period gap (see Figure\,\ref{BPS_PDF_P}).

Regarding the mass transfer rate, compared to the RVJ model, the SBD-optimal model predicts a slightly higher rate at the early stage of the evolution, followed by a lower rate above the period gap. Below the period gap, however, it predicts a higher mass transfer rate. This arises from the differences in MB strength between these two model (the middle panel). 
Below the period gap, MB fully stops in the RVJ model, while it continues with a weaker rate in the SBD-optimal model. It is this residual MB that provides an additional angular-momentum loss below the period gap in the SBD-optimal model, thereby shifting the minimum orbital period to a larger value and yielding better agreement with observations. However, this additional angular momentum loss accelerates the evolution of the system toward longer orbital periods beyond the minimum period. This causes the system to spend a longer time in the period-bouncer phase compared with the RVJ model. For example, the fraction of the total evolutionary timescale spent in the period-bounce phase is 86\% and 67\% for the SBD-optimal and RVJ models, respectively. This suggests that the SBD model may further aggravate the period-bouncer problem (see Section\,\ref{result-BPS}).

\begin{figure*}[ht]
\centering
\includegraphics[width=17cm,height=7cm]{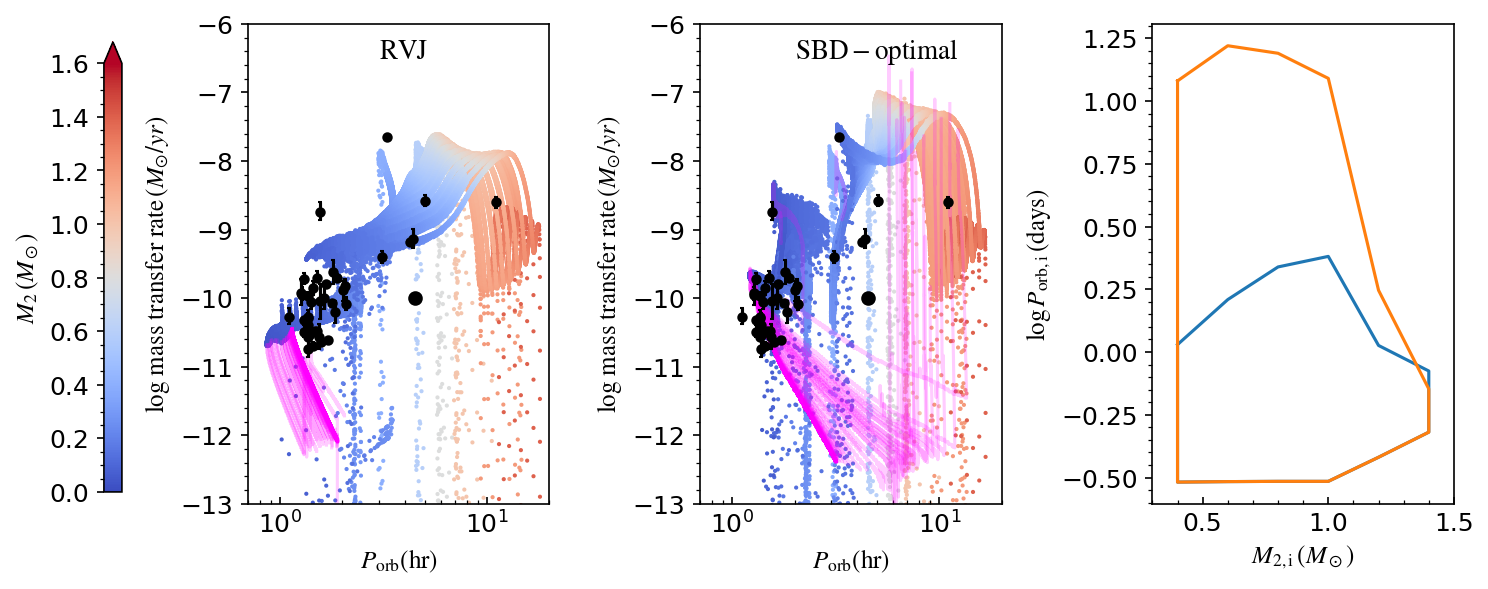}
 \caption{Left and middle panels: a comparison between theoretical and observed mass transfer rates. The theoretical data include all evolutionary tracks with $M_{\rm WD,i}=0.8\,M_{\odot}$ that evolve into CVs in our {\tt MESA} calculations. We show only the case with $M_{\rm WD,i}=0.8\,M_{\odot}$ as a representative example, since the results for other initial WD masses are similar. The left panel represents the RVJ model, while the middle panel is for the SBD-optimal model. In each panel, the magenta lines represent the evolution beyond the minimum orbital period, while the colored dots show the evolution prior to it, with the color bar indicating the companion mass. The black dots are observed sources from \cite{2022MNRAS.510.6110P}. In the middle panel, the sharply increase of mass transfer rates at the end of evolution for some tracks results from that the envelope of the companion is nearly completely stripped, leading to a rapid decrease in $\tau_{\rm conv}$ (but not all of systems show such a state), Right panel: parameter space for the formation of CVs with $M_{\rm WD,i}=0.8\,M_{\odot}$. The blue contour is for the RVJ model, while the yellow contour is for the SBD-optimal model. The two contours coincide along their lower and partially along their side boundaries.}
\label{grid_P-Mdot}
\end{figure*}

 \begin{figure*}[ht]
\centering
\includegraphics[width=14cm,height=7cm]{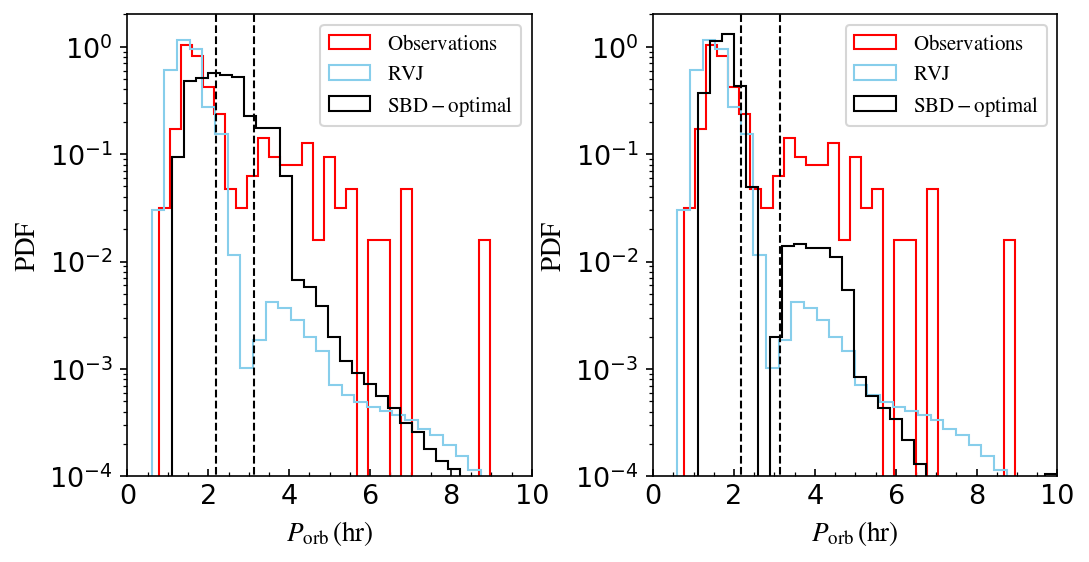}
 \caption{Predicted probability density distributions of orbital period, compared with the observation (red curves). The observed periods for CVs are adopted from \cite{2023MNRAS.524.4867I}.  In each panel, the blue line stands for the RVJ model, while the black line represents the SBD-optimal model. The dashed lines show the period gap. The difference between the left and right panels is that, for the SBD-optimal model, the left panel shows the period distribution considering the entire evolution of CV phase, while the right panel excludes the post–minimum-period evolution with $P_{\rm orb} > 2$ hr.}
\label{BPS_PDF_P}
\end{figure*}

 \begin{figure*}[ht]
\centering
\includegraphics[width=7cm,height=7cm]{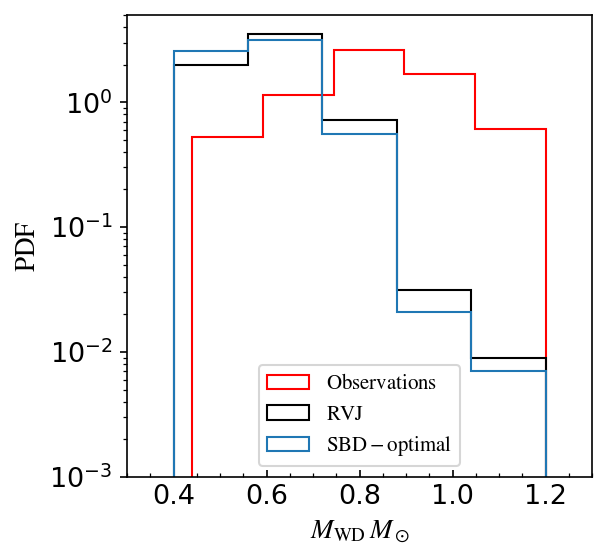}
 \caption{Predicted probability density distributions of WD mass in CVs, compared with the observation (red curves). The observed data are adopted from \cite{2022MNRAS.510.6110P}. The blue line stands for the RVJ model, while black line is for the SBD-optimal model.}
\label{BPS_PDF_WMD}
\end{figure*}

\subsection{Statistical properties of CVs}\label{result-BPS}
 
 In this section, combining the results of  {\tt MESA} with   {\tt BSE}, we investigate the overall properties of the CV population. The method can be briefly summarized as below.
 
  We use MESA to evolve a dense grid of WD+MS systems, with the initial parameter space described in Section \ref{subsection-Basic assumption}. This parameter space is discretized into $N$ bins, where $N = N_1 \times N_2 \times N_3$, and $N_1, N_2, N_3$ represents the number of initial white dwarf masses, donor masses, and orbital periods, respectively. Each bin corresponds to a set of initial parameters ($M_{\rm WD,i}, M_{\rm 2,i}, P_{\rm orb,i}$). For every bin, we perform detailed binary evolution calculations using MESA to obtain the evolutionary tracks of the CV phase.

We then use the Monte Carlo code {\tt BSE} \citep{2000MNRAS.315..543H, 2002MNRAS.329..897H} to simulate the evolution of $10^6$ binaries consisting of two zero-age MS stars with solar metallicities, until the formation of WD+MS binaries. It should be noted that we do not use BSE to further evolve these systems through the CV phase. Instead, for each WD+MS system produced by BSE, we identify the closest bin ($M_{\rm WD,i}, M_{\rm 2,i}, P_{\rm orb,i}$) in the MESA grid and adopt the corresponding MESA evolutionary track to represent its subsequent CV evolution.

 In this way, the WD+MS systems generated by BSE are mapped onto the MESA parameter grid, allowing us to determine the birthrate associated with each MESA bin. By combining these birthrates with the evolutionary tracks obtained from MESA, we finally derive the statistical distribution of CVs.
 
 The key input parameters for {\tt BSE} are similar to \cite{2024ApJ...977...34T}. A constant star formation rate ($5\,M_{\rm \odot}\,\rm yr^{-1}$) is assumed over the past 10\,Gyr. The primary masses are sampled according to initial mass function of \cite{1993MNRAS.262..545K}, and the secondary masses are taken from an evenly distributed mass ratio in $[0,1]$. The initial orbits are assumed to be circular, with the binary separation in logarithm follows a flat distribution within $[3,10^{\rm 4}]\,R_{\rm \sun}$. During mass transfer, we assume that half of mass lost by the donor is accreted by the accretor, as suggested by \citet{2014ApJ...796...37S} due to its better reproduction of Be/X-ray binaries compared with other mass accretion efficiency recipes. For the common envelope evolution, we take the common envelope ejection efficiency $\alpha_{\rm CE}=1/3$ (\citealt{2011A&A...536A..42Z}; \citealt{2022MNRAS.513.3587Z,2023MNRAS.518.3966S}), and the results in \cite{2010ApJ...716..114X} for the binding energy parameter $\lambda$ of the envelope.
 
From MESA, we obtain 1045 and 519 evolutionary tracks that evolve into CVs for the SBD-optimal model and RVJ model, respectively.  As an illustrative example,  the parameter space for $M_{\rm WD,i}=0.8\,M_{\rm \odot}$ is shown in the right panel of Figure\,\ref{grid_P-Mdot}. It is seen that the SBD model can greatly enlarges the parameter space due to its initially stronger MB strength at the early stage of evolution. %at low mass, while it can narrow the parameter space at high mass. 
 %为什么小质量的地方是增加，大质量的地方是减小。分析到底什么情况下RVJ>SBD,什么情况下小于

 A comprehensive comparison with observed mass transfer rates is presented in Figure\,\ref{grid_P-Mdot} (left and middle panels). We find that the RVJ model cannot reproduce the scatter in the distribution of observed mass transfer rate  in both the short and long orbital period regimes. In contrast, while the SBD-optimal model can account for almost all observed systems at short orbital periods, it tends to overestimate the mass transfer rates in the long orbital period regime\footnote{It should be noted that although the SBD-optimal model can explain the two outliers with abnormally high mass transfer rates (or unusually high WD effective temperatures) (SDSS J153817.35 + 512338.0 and DW UMa), their high effective temperatures may reflect special evolutionary stages, such as the onset of mass transfer or a recent nova eruption, during which the white dwarfs have not yet cooled down \citep{2017MNRAS.466.2855P, 2022MNRAS.510.6110P}. }.%A successful magnetic braking model should simultaneously reproduce multiple observational constraints, including the orbital-period distribution and mass transfer rates.}  

The orbital periods of CVs are one of the most well-determined parameters. We compare our results with the observed period distribution in Figure\,\ref{BPS_PDF_P}. The difference between the left and right panels in Figure\,\ref{BPS_PDF_P} is that, for the SBD-optimal model, the left panel shows the period distribution considering the entire evolution of CVs, while the right panel excludes the post–minimum-period evolution with $P_{\rm orb} > 2$ hr. From the left panel, it is found that the RVJ model shows the bimodal distribution in the orbital period, and it exhibits an apparent period gap. It seems that the RVJ model can reproduce the lower boundary of the period gap, while its predicted upper boundary slightly larger than the observed one. Moreover, it underestimates the number of long-period CVs.

However, the SBD-optimal model shows only a single peak in the distribution of orbital period. This occurs because the residual MB accelerates the evolution of CVs toward longer orbital periods beyond the minimum period, allowing systems to cross the period gap again and thereby increasing the number of systems within the gap. Meanwhile, this residual MB further aggravates the period bouncer problem, as the model predicts a period-bouncer fraction exceeding 80\%, compared to $\sim 70\%$ predicted by the standard model \citep[see also e.g. ][]{1993A&A...271..149K,2018MNRAS.478.5626B} and the observed value of $31 \pm 7\%$ \citep{2020MNRAS.494.3799P}, or even only a few percent as suggested by \cite{2023MNRAS.525.3597I}. Instead, if the post–minimum-period evolution with $P_{\rm orb} > 2$ hr is excluded in the SBD-optimal model, the period gap will become obvious, as shown by black curve in the right panel of Figure\,\ref{BPS_PDF_P}. If so, the SBD-optimal model better reproduces the period gap than the RVJ model. This may imply that our current understanding for the SBD model is still incomplete. For the SBD model to properly describe the evolution of CVs, it is necessary to ensure that the predicted CV evolutionary tracks do not cross the period gap again after reaching the minimum period. Even so, the SBD-optimal model still faces some difficulties. For example, as shown in the right panel of Figure\,\ref{BPS_PDF_P}, it significantly underestimates the number of systems with orbital periods $\gtrsim 5$\,hr. In addition, the observed period bouncers are rare, and there appears to be a cutoff, with few or no systems found below a mass transfer rate of $\log \dot M_{2}\lesssim -11$, as shown in the right panel of Figure\,\ref{Fig-example1}.  However, the SBD-optimal model predicts a large number of such systems. We emphasis that the accurate result for the SBD model is also dependent on the prescription of $\tau_{\rm conv}$, see Section \ref{Effect-tau}.

Figure \ref{BPS_PDF_WMD} displays the predicted WD mass distribution in comparison with observations. We find that the WD mass distributions predicted by the RVJ and SBD models do not show significant differences. In both models, the WD mass is mainly concentrated in the range of $0.4-0.6\,M_{\rm \odot}$, which is consistent with previous work \citep[e.g. ][]{2016MNRAS.455L..16S}. However, the observed WD masses are significantly higher than predicted, with most of them in the range of $0.8-1.2\,M_{\rm \odot}$ \citep[e.g. ][]{2011A&A...536A..42Z, 2022MNRAS.510.6110P}. Thus, we conclude that the SBD model cannot solve the WD mass problem.

\section{Discussion}\label{Discuss}
In the above sections, we have investigated the effect of four MB models on the evolution of CVs. In what follows we discuss some uncertainties and the implications of our results.

\subsection{Comparison with recent works}\label{Discus-comparison}
During the submission and review of this manuscript, some recent studies have conducted similar investigations, including \cite{2026A&A...707A..76Z} and \cite{2026arXiv260314560B}. We briefly compare our work with them.

\cite{2026A&A...707A..76Z} investigated the effect of five MBs on the CV evolution, including the RVJ model,  Matt model \citep{2012ApJ...754L..26M}, RM12 model \citep{2012ApJ...746...43R}, ‘Intermediate’ model and $\tau$-boosted model \citep{2019MNRAS.483.5595V}. In addition to the RVJ model, the model that is commonly adopted in both our work and theirs is the $\tau$-boosted model. Therefore, our work complements their study by exploring additional MBs, and provides a more comprehensive view of their impact on CV evolution. In the following, we compare the results of the $\tau$-boosted model adopted in both studies.

The example shown in Figure\,\ref{Fig-example1} has similar initial parameters to those in the upper panel of Figure\,5 in \cite{2026A&A...707A..76Z}. The only difference is that our initial orbital period is 0.339\,days, while theirs is 0.4\,days. We also evolve a system with an initial orbital period of 0.4\,days for our $\tau$-boosted model and find that the results are nearly identical to those obtained with an initial orbital period of 0.339\,days. Therefore, we can directly compare the evolutionary track of the $\tau$-boosted model shown in our Figure\,\ref{Fig-example1} with the results of \cite{2026A&A...707A..76Z}.

The evolutionary trends in our track and theirs are generally similar: above the period gap, both tracks show an trend of increase in orbital period after the onset of mass transfer. However, the upper boundary of the period gap differs, being $\sim 4.7$\,hr in their result and $\sim 5.5$\,hr in ours. This difference is likely caused by the treatment of the convective turnover timescale $\tau_{\rm conv}$, which leads to different MB strengths. As shown in their Figure\,7, the MB strength remains below $\sim 10^{36}$\,[dyn\,cm] throughout almost the entire evolution, including the mass transfer phase, whereas in our model it is generally larger than $10^{36}$\,[dyn\,cm] (middle panel of Figure\,\ref{Fig-example1}). The stronger MB in our model leads to a larger upper boundary of the period gap,  because the higher mass-transfer rate drives the donor further out of thermal equilibrium, causing it to maintain a more inflated radius.

Although \cite{2026A&A...707A..76Z} did not evolve a dense grid of initial parameters, they have already explored CV evolution over a wide range of initial parameters through several representative examples. They varied the initial WD mass from $0.6$ to $1.0\,M_{\rm \odot}$ and the initial companion mass from 0.6 to 1.2\,$M_{\rm \odot}$. Their results show that only tracks with a initial massive WD ($>1.0\,M_{\rm \odot}$) and a initial massive companion ($\gtrsim1.2\,M_{\rm \odot}$) can produce an upper boundary of the period gap around $\sim 3$\,hr, roughly consistent with the observed value ($\sim 3.18$\,hr). For other initial parameters, the upper boundary is significantly higher than observed. Therefore, we can safely conclude that the $\tau$-boosted model in our work cannot reproduce the observed period gap of CVs. This is because the MB strength in our model is stronger than that in \cite{2026A&A...707A..76Z}, which results in a higher upper boundary. Moreover, from a population synthesis perspective, although systems with an initial massive WD and an initial massive companion have an upper boundary similar to observed, they do not dominate CV population.

\begin{figure*}[ht]
\centering
\includegraphics[width=8cm,height=8cm]{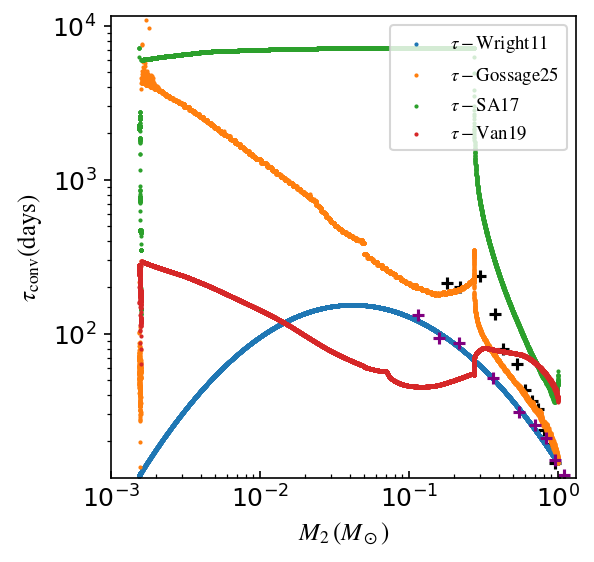}
 \caption{ Comparison of $\tau_{\rm conv}$ adopted from different literature. The orange dots show the theoretical $\tau_{\rm conv}$ from the Gossage25 prescription (Eq.\,(\ref{Eq-tau-Gossa})), which is used in this work. The red dots represent the values from the Van19 prescription (Eq.\,(\ref{Eq-tau-Van19})). The green dots indicate the values from the SA17 prescription (Eq.\,(\ref{Eq-tau-XuXT})), which are used in study of the $\tau$-boosted and CARB models in \cite{2021ApJ...909..174D}. The blue dots are the empirical $\tau_{\rm conv}$ from the Wright11 prescription (Eq.\,(\ref{Eq-tau-conv})), which is utilized in study of the SBD model in \cite{2025A&A...696A..92B}. The above data are obtained by evolving a binary system. The black and purple crosses denote the observed $\tau_{\rm conv}$ from \cite{2025ApJ...988..102G} and \cite{2011ApJ...743...48W}, respectively. Note that these observational data are only available for stars with masses $\gtrsim 0.1\,M_{\rm \odot}$.}
\label{Fig-tau-conv}
\end{figure*}

\cite{2026arXiv260314560B} presented a revised SBD model. Different from our work and their previous study \citep{2025A&A...696A..92B}, which adopted the saturated MB  developed by \cite{1995ApJ...441..865C}, their new study employs the saturated MB formulation suggested by \cite{2015ApJ...799L..23M}. 
%The main difference between these two saturated MB prescriptions lies in the different power law dependences of the MB strength on the donor radius, donor mass, and convective turnover timescale. 
Importantly, instead of adopting an empirical $\tau_{\rm conv}$ expression fitted from observations used in their previous work \citep{2025A&A...696A..92B}, they also use an explicit method to calculate the convective turnover timescale. According to their Figure 1, the $\tau_{\rm conv}$ value exhibits a bump around $0.3$–$0.4\,M_{\rm \odot}$, consistent with the trend reported by \cite{2025ApJ...988..102G}, which is adopted in this study. However, a comparison shows that their calculated $\tau_{\rm conv}$ values are slightly larger than ours; for example, at $M = 0.1\,M_{\rm \odot}$, their $\tau_{\rm conv}$ is about 350\,days, whereas ours is about 200\,days (Figure\,\ref{Fig-tau-conv}). Based on new settings , they derive an optimal parameter for the SBD model to be $K = 20$ and $\eta \approx 2-3$, while our optimal solution is $K \approx 140$ and $\eta \approx 20$. Similar to our findings, they also show that a constant $\eta$ can reproduce almost all observed mass transfer rates below the period gap, whereas above the period gap the predicted mass transfer rates are higher than the observed values.

 \begin{figure*}[ht]
\centering
\includegraphics[width=14cm,height=10cm]{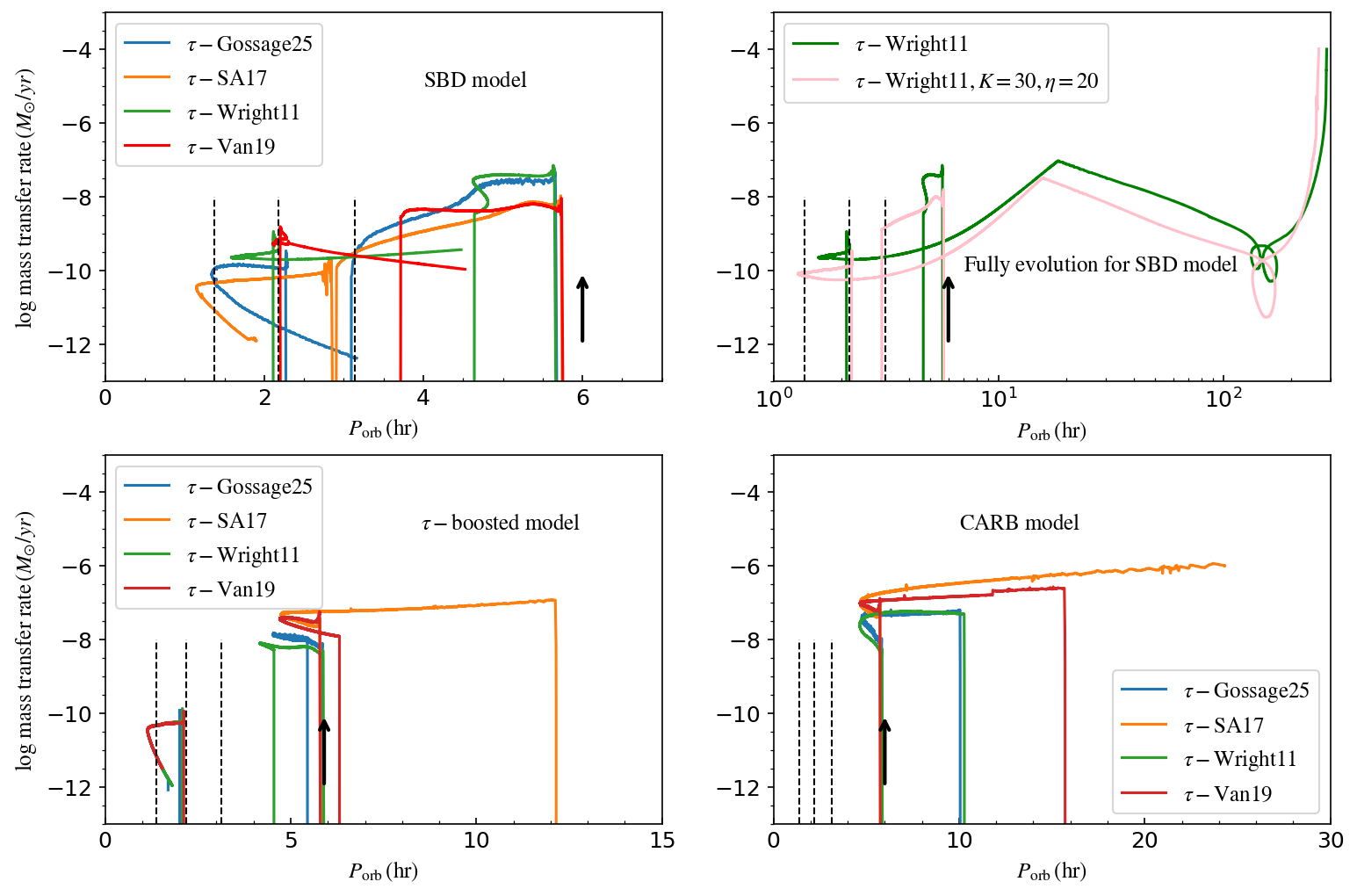}
 \caption{The effect of different $\tau_{\rm conv}$ prescriptions on CV evolution under the SBD model (upper panels), $\tau$-boosted model (lower left), and CARB model (lower right). All tracks are obtained by evolving a binary with initial parameters of $M_{\rm WD,i}=0.8\,M_{\rm \odot}$, $M_{\rm 2,i}=0.8\,M_{\rm \odot}$, and $P_{\rm orb,i}=0.6\,\rm days$. In each panel, the three dashed lines from left to right indicate the minimum orbital period, and the lower and upper boundaries of the period gap, respectively. The black arrows have the meaning same as Figure\,\ref{Fig-example1}.
In the three panels other than the upper right one, different colored lines represent evolutionary tracks adopting different $\tau_{\rm conv}$ prescriptions (see the legend). In the upper left panel, we adopt $K = 140$ and $\eta = 20$. For the Van19 and Wright11 prescriptions, only the evolutionary phase with $M_{2} > 0.01\,M_{\rm \odot}$ is shown. The full evolutionary track of the Wright11 prescription, including the phase with $M_{2} < 0.01\,M_{\rm \odot}$, is presented in the upper right panel. In addition, the upper right panel also shows an example of the SBD model with $K=30$ and $\eta=20$ (pink line), which can best reproduce the minimum orbital period and the period gap of CVs in the Wright11 prescription for $\tau_{\rm conv}$.}
%\textbf{Note that for the Wright11 model in the SBD prescription, the calculation terminates due to an excessively small timestep caused by an extremely high mass transfer rate.}}
\label{Fig-tau-effect}
\end{figure*}

Overall, although the detailed expression for the saturated MB and the treatment of $\tau_{\rm conv}$ are slightly different, our SBD-optimal model yields CV evolutionary tracks that are very similar to those predicted by their optimal model. The main difference lies in the scope of the study: we perform a population synthesis calculation, whereas they present only several representative evolutionary examples. Given the high similarity of the evolutionary tracks, it is expected that their model would have similar predictions and encounter similar difficulties as those revealed by our results, such as predicting a large number of systems remaining in the period gap after passing the minimum orbital period \footnote{In their Figure 2, the evolution does not cross the period gap again during the period-bouncer phase. We suspect that this is because their calculations are artificially terminated earlier and the systems are not evolved to a Hubble timescale.} and a more severe overestimation of the fraction of the period bouncer compared to the standard model.

\subsection{A residual magnetic braking below the period gap?}

MB originates from the coupling between the companion’s magnetic field and its stellar winds. Under the traditional assumption, MB ceases abruptly once the companion becomes fully convective, because the conventional viewpoint holds that fully convective stars do not possess magnetic fields. In contrast, in the SBD model, a residual MB operates below the period gap, although this residual MB is one order of magnitude weaker than that above the gap. This contradicts the standard assumption. Nevertheless, several observations suggest that fully convective stars can in fact maintain strong magnetic fields \citep{2016Natur.535..526W, 2022A&A...662A..41R}. Therefore, it remains plausible that a residual MB operates below the period gap \citep[see][for a more detailed discussion]{2011ApJS..194...28K}.

In our {\tt MESA} calculations, the systems can evolve down to donor masses below $\sim 10^{-3}\,M_{\rm \odot}$ for the SBD-optimal model, in which residual MB operates throughout the evolution. However, our understanding of magnetic-field generation and the conditions required for MB in brown dwarfs remains very limited. At present, the available observational constraints on $\tau_{\rm conv}$ are mainly restricted to stars with masses $>0.1\,M_{\rm \odot}$ \citep{2011ApJ...743...48W, 2025ApJ...988..102G}. Below this mass, $\tau_{\rm conv}$ is obtained through explicit calculations in our study, but lacks observational constraints. Therefore, better observational constraints on $\tau_{\rm conv}$ for low-mass stars would significantly improve theoretical models. %it remains to be investigate whether the same MB prescription can be applied to very low-mass stars and sub-brown-dwarf objects.

\subsection{The effect of different $\tau_{\rm conv}$ prescriptions}\label{Effect-tau}

Currently, there exist several prescriptions for the convective turnover timescale $\tau_{\rm conv}$ in the literature \citep{2019MNRAS.483.5595V, 2011ApJ...743...48W, 2017MNRAS.472.2590S, 2025ApJ...988..102G}. Since the $\tau$-boosted, CARB, and SBD MB models all depend on $\tau_{\rm conv}$, it is important to explore how different $\tau_{\rm conv}$ prescriptions affect their predictions. Here, we collect four commonly used $\tau_{\rm conv}$ prescriptions and investigate their impact on CV evolution under these three MB models. The prescription of \cite{2025ApJ...988..102G} (Gossage25) has been given in Section\,\ref{subsection-MBs}. The remaining three prescriptions are listed below.

(1) Van19 model: \cite{2019MNRAS.483.5595V} calculated $\tau_{\rm conv}$ as follow % (Their code is available on \url{https://zenodo.org/records/2592801}),}
\begin{equation}\label{Eq-tau-Van19}
\tau_{\text{conv}} = \int_{R}^{R_{\rm s}} \, \frac{d {r}}{v_{\rm conv}},
\end{equation}
$R$ and $R_{\rm s}$ are the bottom and the top of the outer convective zone, respectively, while $v_{\rm conv}$ is the local convective velocity.

(2) SA17 model: In \cite{2021ApJ...909..174D}, they adopted the prescription of $\tau_{\rm conv}$ suggested by \cite{2017MNRAS.472.2590S} to investigate the effect of different MBs on the evolution of NS binaries. This prescription depends on the mass of the external convection zone $M_{\rm cz}$,
 \begin{eqnarray}\label{Eq-tau-XuXT}
\log_{10}(\frac{\tau_{\rm conv}}{\rm 1\,s}) 
= & 8.79 - 2|\log_{10}(m_{cz})|^{0.349} - 0.0194|\log_{10}(m_{cz})|^2
 \nonumber \\
 & - 1.62 \min[\log_{10}(m_{cz}) + 8.55, 0],
\end{eqnarray}
where $m_{\rm cz} = M_{\rm cz}/M_{\rm 2}$.

(3) Wright11 model: To investigate the evolution of CVs under the SBD model, \cite{2025A&A...696A..92B} employed the empirical $\tau_{\rm conv}$ obtained by fitting the observed results of \cite{2011ApJ...743...48W}. This empirical formulation is a function of companion mass,
\begin{equation}\label{Eq-tau-conv}
\log_{10} \left( \frac{\tau_{\rm conv}}{\mathrm{d}} \right) = 1.16 - 1.49 \log_{10} \left( \frac{M_2} {M_{\odot}} \right) - 0.54 \log_{10}^2 \left( \frac{M_2}{M_{\odot}} \right).
\end{equation}

Figure\,\ref{Fig-tau-conv} presents a comparison of the four $\tau_{\rm conv}$ prescriptions. The theoretical data are obtained by implementing the above equations into the {\tt MESA} code and evolving a binary system numerically. A comparison of the observational data from \cite{2011ApJ...743...48W} and \cite{2025ApJ...988..102G} shows that the two data are generally consistent at the high-mass end. However, at the low-mass end, the observed values from \cite{2025ApJ...988..102G} are slightly higher than those from \cite{2011ApJ...743...48W}. At a mass of $0.3\text{–}0.4\,M_{\odot}$, the data from \cite{2025ApJ...988..102G}  exhibits a noticeable bump, whereas the \cite{2011ApJ...743...48W} results show a smooth variation. \cite{2011ApJ...743...48W} adopted a quadratic function in logarithmic space to fit their observational data (Eq.\,(\ref{Eq-tau-conv})), resulting in a decrease of $\tau_{\rm conv}$ with decreasing mass at the low-mass end. This behavior differs from the other three $\tau_{\rm conv}$ prescriptions.

Comparing the theoretical $\tau_{\rm conv}$, the SA17 prescription  gives the highest values, which are also larger than any observational data. Although the Van19 prescription calculates $\tau_{\rm conv}$ using an explicit method, its results are not consistent with the observations. In principle, except for the Wright11 precription, the calculation for other three prescriptions are all related to the size of the outer convective zone. Therefore, these three prescriptions show a sharp decrease in $\tau_{\rm conv}$ at the very low-mass end ($\sim 10^{-3}\,M_{\odot}$) because of the ultra-stripping of the outer envelope at the late stage of evolution. This sharp decrease in $\tau_{\rm conv}$ results in the increase of MB strength in the SBD model,  because $\dot J_{\rm MB, SBD} \propto \left(\frac{1}{\tau_{\rm conv}}\right)^2$. This can explain the sharp increase in the mass transfer rate at the end of the evolution for some tracks under the SBD model shown in the middle panel of Figure\,\ref{grid_P-Mdot}.

\begin{figure*}[ht]
\centering
\includegraphics[width=16cm,height=7cm]{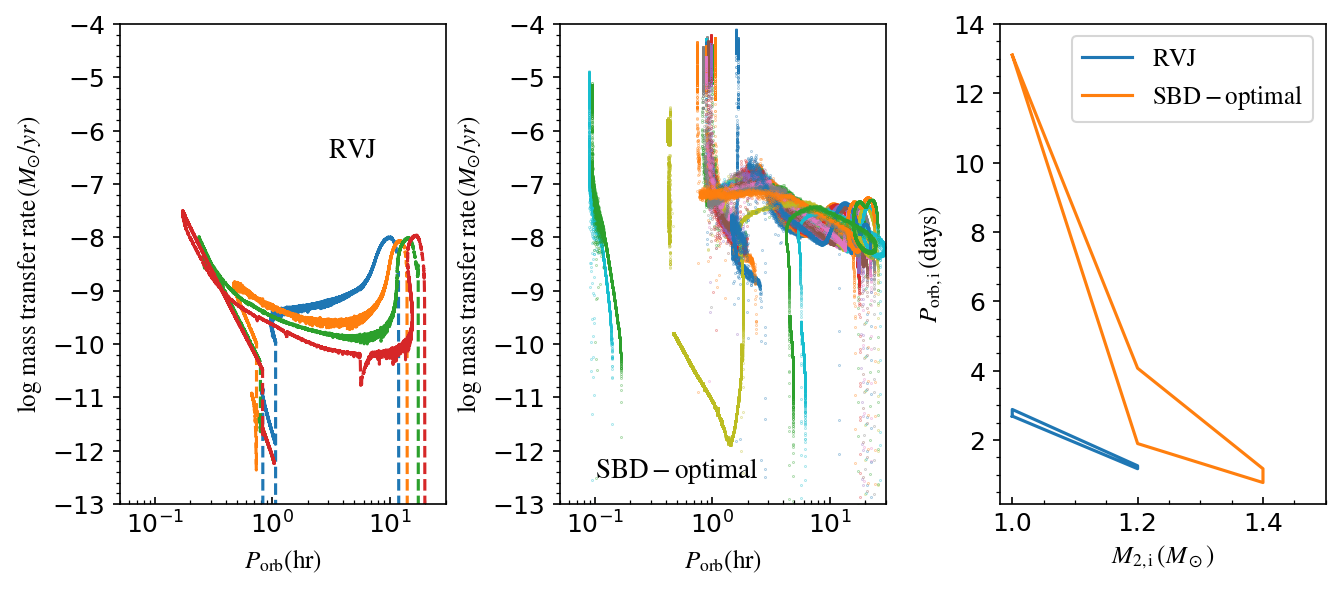}
 \caption{Comparison of the formation and evolution of AM CVn systems between the RVJ model and the SBD-optimal model. The left and middle panels show the evolution of the mass transfer rate as a function of orbital period for the RVJ model and SBD-optimal model, respectively. The evolution proceeds from right to left. Each line represents an evolutionary track that leads to the formation of an AM CVn system. The right panel shows the initial parameter space that can produce AM CVn systems for a representative case with $M_{\rm WD,i}=0.8\,M_{\rm \odot}$. The blue and orange lines represent the RVJ and SBD-optimal models, respectively.}
\label{Fig-AMCVn}
\end{figure*}

In Figure\,\ref{Fig-tau-effect}, we investigates the effect of different $\tau_{\rm conv}$ prescriptions on CV evolution under the SBD, $\tau$-boosted, and CARB models. We first focus on the $\tau$-boosted and CARB models. In the CARB model (lower right panel), the orbit expands from $\sim 5.8\,\rm hr$ to $>10\,\rm hr$ during the mass transfer phase for all $\tau_{\rm conv}$ prescriptions. Then MB stops and system enters detached phase under the Gossage25, Wright11, and Van19 prescriptions. The companion cannot fill its Roche lobe again after MB ceases for the evolution under the three $\tau_{\rm conv}$ prescriptions, even if the system evolves up to a Hubble timescale. In contrast, under the SA17 prescription, MB continues to operate until the companion mass decreases to $\sim 0.06\,M_{\rm \odot}$, after which the calculation becomes non-convergent due to an excessively small timestep. In the $\tau$-boosted model (lower left panel), the evolution under the SA17 model is similar to that in the CARB model. For the other three $\tau_{\rm conv}$ prescriptions, the systems can still experience mass transfer after MB ceases, but none of them can reproduce the orbital properties of CVs. In both the $\tau$-boosted and CARB models, the mass transfer rates predicted by the SA17 and Van19 prescriptions are higher than those of the other two prescriptions because they yield larger $\tau_{\rm conv}$ values (at high mass end for the SA17 prescription), and $\dot{J}_{\rm MB} \propto (\tau_{\rm conv})^{\gamma}$ with $\gamma>0$ (Eqs.\,(\ref{Eq-tau-boosted}) and (\ref{Eq-CARB})).

Clearly, neither the CARB nor the $\tau$-boosted model can reproduce the orbital properties of CVs regardless of the adopted $\tau_{\rm conv}$ prescription. It is worth noting that these MB models contain no free parameters. This may imply that they are not suitable for CV evolution. Our result also highlights that, when applying these two MB models to neutron star or black hole binaries and drawing conclusions about their applicability, the effect of $\tau_{\rm conv}$ must be carefully taken into account, particularly when interpreting individual systems.

We next concentrate on the SBD model. Since the MB strength in the SBD model $\propto \left (\frac{1}{\tau_{\rm conv}}\right)^{2}$, it predicts a lower mass transfer rate at long orbital periods when adopting the SA17 and Van19 prescriptions for $\tau_{\rm conv}$. This is in contrast to the $\tau$-boosted model and the CARB model. The upper boundary of the period gap is larger than observed in the Van19 and Wright11 prescriptions, while almost no period gap is formed in the SA17 prescription. Except for the SA17 prescription, the other three prescriptions can roughly reproduce the lower boundary of the period gap, whereas only the Gossage25 prescription can produce a minimum orbital period consistent with observations. We note, however, that adopting different values of $K$ and $\eta$ can change these results. Overall, this further emphasizes the importance of accurately determining $\tau_{\rm conv}$ in these MB models.

Among the results obtained with the SBD model, the case adopting the $\tau_{\rm conv}$ prescription of Wright11 is particularly interesting. This configuration was also used in \cite{2025A&A...696A..92B}. Note that \cite{2025A&A...696A..92B} halted their simulations once the donor star reached a mass of $0.05\,M_{\rm \odot}$ (A. Barraza-Jorquera, private communication), it is interesting to explore the further evolution of the system below $M_{2}<0.05\,M_{\rm \odot}$ in order to investigate its ultimate fate. This is examined in the upper right panel of Figure\,\ref{Fig-tau-effect}.%In the upper right panel of Figure\,\ref{Fig-tau-effect},  only the track with $M_{2} >0.01\,M_{\rm \odot}$ is shown for the model with Wright11 prescription, while its full evolutionary track is presented in the upper right panel of Figure\,\ref{Fig-tau-effect}. In addition, in the upper left panel of Figure\,\ref{Fig-tau-effect}, we also present another example with $K=30$ and $\eta=20$ (pink line), which best reproduces the minimum orbital period and the period gap of CVs in the Wright11 $\tau_{\rm conv}$ prescription. These cases exhibit interesting evolutionary behavior.}

Under the effect of a residual MB below the period gap, the system evolves toward increasing orbital period and mass transfer rate beyond the minimum orbital period. Then, when reaching a critical point at $P_{\rm orb}\sim 15-20\,\rm hr$, the mass transfer rate declines. This is because the system transitions from the saturated regime to the unsaturated regime, leading to a change in the angular momentum loss rate. When the orbit expands to $\sim 200\,\rm hr$, the mass transfer rate reaches a local minimum value \footnote{The loop at $\sim 200\,\rm hr$ arises from non-smooth variations in the companion's radius predicted by the `tau\_100\_tables' atmosphere model at companion mass $\sim 10^{-4}\,M_{\rm \odot}$. In our {\tt MESA} code, when the companion mass decreases to $0.08\,M_{\rm \odot}$, we change `which\_atm\_option' from the default value to `tau\_100\_tables'. This table is primarily designed for the evolution of low-mass stars, brown dwarfs, and planets \citep{2011ApJS..192....3P}.}. Then, the mass transfer rate sharply increases to $>10^{\rm -4} M_{\rm \odot}\,\rm yr^{-1}$ and the calculation stops due to too small a timestep (The companion has a final mass of about $1\times 10^{\rm -4}\,M_{\rm \odot}$ and a logarithmic age (in years) of 9.8). The sharply increase of mass transfer rate can be attributed to the combination of low companion mass and high $\dot J_{\rm MB}$ arising from small $\tau_{\rm conv}$ obtained by Wright11 prescription (Figure\,\ref{Fig-tau-conv}). This indicates that the companion may be destroyed through merging or tidal disruption. This appears to align with the recent suggestion by \cite{2023MNRAS.525.3597I}. They found that period bouncers constitute only a few percent of the total CV population and proposed that, in the final phase of their evolution, CV may destroy its donor. \cite{1998PASP..110.1132P} also suggested that it is probably necessary to have some means of destroying CVs before they reach the predicted very high space densities. For the Wright11 prescription with $K=30$ and $\eta =20$, the total evolutionary timescale is $3.11\times10^{9}\,\rm yr$, while the period-bouncer phase lasts for $\sim9.8\times10^{8}\,\rm yr$, corresponding to about 31\% of the total evolutionary timescale . This means that it may effectively reduce the fraction of the period bouncer compared with the RVJ model ($\sim 70\%$). The evolution of period bouncers does not proceed to very low mass transfer rates, which appears to be consistent with the lack of period bouncers with low mass transfer rate found by \cite{2022MNRAS.510.6110P} (the right panel of Figure\,\ref{Fig-example1}).

However, we argue that whether the final disruption predicted by the above model is realistic depends critically on whether the Wright11 prescription can accurately describe $\tau_{\rm conv}$ in low-mass stars. Unfortunately, this issue remains poorly understood due to the lack of observational constraints on $\tau_{\rm conv}$ in low-mass stars. In particular, the SBD model combined with the Wright11 prescription predicts a significant population of period bouncer systems with low masses and high mass transfer rates. For example, for the tracks shown in the upper right panel of Figure\,\ref{Fig-tau-effect}, the fraction of the total evolutionary timescale spent in period bouncer phase with $P_{\rm orb}>2\,\rm hr$ is $\sim 10\%$. However, such systems have not yet been identified in significant numbers in observations. 

\subsection{Implementing the SBD-optimal model to AM CVn}\label{AM CVn}

 As is well known, the RVJ model leads to a fine-tuning problem in the formation of ultra-compact systems, such as AM CVns and binaries with ELM WDs ($M_{\rm WD}\lesssim 0.2\,M_{\odot}$). For these systems, only a very narrow range of initial parameters (perhaps the range of initial orbital periods $\Delta P_{\rm orb}$ is only several hours) can lead to their formation under the RVJ model \citep{2014A&A...571A..45I}. \cite{2023A&A...678A..34B} and \cite{2021MNRAS.506.3266S} demonstrated that the CARB model can resolve the fine-tuning issue.

In this section, we discuss the influence of the SBD-optimal model on the formation and evolution of AM CVn systems. The results are shown in Figure\,\ref{Fig-AMCVn}. We identify an initial system to be capable of evolving into an AM CVn-like system if it's evolution is convergent (i.e., toward shorter periods) and the mass-transferring donor develops a degenerate helium core. The right panel indicates that the SBD-optimal model enlarges the parameter space for AM CVn formation due to its initially stronger MB strength. However, the mass transfer rates predicted by the SBD-optimal model exhibit distinct behavior compared with the RVJ model. First, the SBD-optimal model predicts higher mass transfer rates at long orbital periods. Second, most systems reach extremely high mass transfer rates ($\sim10^{-4}\,M_{\rm \odot}\,\rm yr^{-1}$) when the orbital period decreases to $\sim1\,\rm hr$, after which the calculations terminate due to an excessively small timestep. Only a few tracks can evolved to $P_{\rm orb}<60\,\rm min$. In contrast, most of the observed AM CVn have orbital periods $\lesssim 60\,\rm min$ \citep{2018A&A...620A.141R, 2023JAVSO..51..227B, 2024RNAAS...8..299M}. Whether the prediction of the SBD-optimal model is consistent with observations of AM CVn (e.g. the birth rate, number, and element abundance) requires a more comprehensive research, which is beyond the scope of this work.

\section{Summary}\label{Sect-Sum}

Magnetic braking (MB) plays a crucial role in the evolution of close binaries. Increasing observational evidence indicates that the widely used RVJ model is inconsistent with certain observations. Several alternative MB models have been proposed and extensively discussed in the context of X-ray binary evolution; however, they have not yet been systematically tested in CVs. In this work, we investigate the effect of four different MBs on CV evolution. We summarize our results as follows.

(1) We find that both the $\tau$-boosted and CARB models are so strong that they fail to reproduce the location of the period gap, regardless of the adopted $\tau_{\rm conv}$ prescription. Therefore, we argue that the $\tau$-boosted and CARB models are not suitable to CVs.

(2) We perform a detailed comparison between the standard RVJ model and the saturated, boosted, and disrupted (SBD) model. Consistent with previous studies, we find that the RVJ model slightly overestimates the upper boundary of the period gap and it predicts an excessively high fraction of period bouncers ($\sim 70$\%). In addition, it also cannot  explain the scatter in the distribution of observed mass transfer rate distribution in both the short and long orbital period CVs.

(3) For the SBD-optimal model, it can successfully explain the location of the period gap, the minimum orbital period, and the observed mass transfer rates below the period gap. However, it also faces several challenges. For example, it predicts higher mass transfer rates above the period gap compared with observations. In addition, it further aggravates the period bouncer problem, as the model predicts a fraction of the period bouncer exceeding 80\%, compared with $\sim70\%$ predicted by the RVJ model and the observed value of $31 \pm 7\%$ \citep{2020MNRAS.494.3799P}, or even only a few percent \citep{2023MNRAS.525.3597I}. These results highlight that, although the period gap and minimum orbital period offer important diagnostics for testing MB theory, they are not sufficient to fully assess or validate a given model.

(4) We also find that different $\tau_{\rm conv}$ prescriptions have a significant influence on CV evolution in the $\tau$-boosted, CARB, and SBD models (Section\,\ref{Effect-tau}). Depending on the adopted $\tau_{\rm conv}$ prescription, the predicted mass transfer rate, the location of the minimum orbital period and the period gap may differ significantly. In some cases, the period gap may even disappear.  This highlights the importance of accurately determining  $\tau_{\rm conv}$.

(5) Finally, we investigate the impact of the SBD-optimal model on the formation and evolution of AM CVn systems. Although the model can widen the initial parameter space for AM CVn formation, it requires a more detailed study to compare with the observation (see Section~\ref{AM CVn}).

Our results indicate that a comprehensive and unified understanding of MB has yet to be achieved.

\nolinenumbers 
\section*{Acknowledgments}
\quad We are grateful to the anonymous referee for valuable comments and Diogo Belloni's meaningful discussion. T.W.S is supported by National Natural Science Foundation of China under grant No.12503038. L.X.D is supported by the National Key Research and Development Program of China (2021YFA0718500), the Natural Science Foundation of China under grant No. 12041301 and 12121003. C.Z is supported by Shandong Provincial Natural Science Foundation of China (Grant No. ZR2025QC1503). 

\section*{Data Availability}
\quad The MESA code for this work are publicly available at \dataset[doi:10.5281/zenodo.19657109]{https://doi.org/10.5281/zenodo.19657109}.


\begin{thebibliography}{plain}\label{thebibliography}
\bibitem[Andronov et al.(2003)]{2003ApJ...582..358A} Andronov, N., Pinsonneault, M., \& Sills, A.\ 2003, \apj, 582, 1, 358. doi:10.1086/343030
\bibitem[Barraza-Jorquera et al.(2025)]{2025A&A...696A..92B} Barraza-Jorquera, J.~A., Schreiber, M.~R., \& Belloni, D.\ 2025, \aap, 696, A92. doi:10.1051/0004-6361/202553757
\bibitem[Barraza-Jorquera et al.(2026)]{2026arXiv260314560B} Barraza-Jorquera, J.~A., Schreiber, M.~R., Littlefair, S., et al.\ 2026, arXiv:2603.14560. doi:10.48550/arXiv.2603.14560
\bibitem[Begari \& Maccarone(2023)]{2023JAVSO..51..227B} Begari, T. \& Maccarone, T.~J.\ 2023, \jaavso, 51, 2, 227. doi:10.48550/arXiv.2312.06007
\bibitem[Belloni et al.(2018)]{2018MNRAS.478.5626B} Belloni, D., Schreiber, M.~R., Zorotovic, M., et al.\ 2018, \mnras, 478, 4, 5626. doi:10.1093/mnras/sty1421
\bibitem[Belloni \& Schreiber(2023)]{2023A&A...678A..34B} Belloni, D. \& Schreiber, M.~R.\ 2023, \aap, 678, A34. doi:10.1051/0004-6361/202347047
\bibitem[Belloni et al.(2024)]{2024A&A...682A..33B} Belloni, D., Schreiber, M.~R., Moe, M., et al.\ 2024, \aap, 682, A33. doi:10.1051/0004-6361/202347931
\bibitem[Belloni et al.(2025)]{2025A&A...697A.100B} Belloni, D., Schreiber, M.~R., \& El-Badry, K.\ 2025, \aap, 697, A100. doi:10.1051/0004-6361/202553937
\bibitem[Chaboyer et al.(1995)]{1995ApJ...441..865C} Chaboyer, B., Demarque, P., \& Pinsonneault, M.~H.\ 1995, \apj, 441, 865. doi:10.1086/175408
\bibitem[Chen et al.(2021)]{2021MNRAS.503.3540C} Chen, H.-L., Tauris, T.~M., Han, Z., et al.\ 2021, \mnras, 503, 3, 3540. doi:10.1093/mnras/stab670
\bibitem[Deng et al.(2021)]{2021ApJ...909..174D} Deng, Z.-L., Li, X.-D., Gao, Z.-F., et al.\ 2021, \apj, 909, 2, 174. doi:10.3847/1538-4357/abe0b2
\bibitem[Deng \& Li(2024)]{2024ApJ...971...54D} Deng, Z.-L. \& Li, X.-D.\ 2024, \apj, 971, 1, 54. doi:10.3847/1538-4357/ad5fec
%\bibitem[Deng \& Li(2024)]{2024ApJ...971...54D} Deng, Z.-L. \& Li, X.-D.\ 2024, \apj, 971, 1, 54. doi:10.3847/1538-4357/ad5fec
\bibitem[Echeveste et al.(2024)]{2024MNRAS.530.4277E} Echeveste, M., Novarino, M.~L., Benvenuto, O.~G., et al.\ 2024, \mnras, 530, 4, 4277. doi:10.1093/mnras/stae1115
\bibitem[El-Badry et al.(2022)]{2022MNRAS.517.4916E} El-Badry, K., Conroy, C., Fuller, J., et al.\ 2022, \mnras, 517, 4, 4916. doi:10.1093/mnras/stac2945
\bibitem[Fan et al.(2024)]{2024ApJ...976..210F} Fan, Y.-N., Shao, Y., \& Chen, W.-C.\ 2024, \apj, 976, 2, 210. doi:10.3847/1538-4357/ad8b48
\bibitem[Gossage et al.(2025)]{2025ApJ...988..102G} Gossage, S., Kiman, R., Monsch, K., et al.\ 2025, \apj, 988, 1, 102. doi:10.3847/1538-4357/adde4d
\bibitem[Green et al.(2025)]{2025A&A...700A.107G} Green, M.~J., van Roestel, J., \& Wong, T.~L.~S.\ 2025, \aap, 700, A107. doi:10.1051/0004-6361/202554925
\bibitem[Hurley et al.(2002)]{2002MNRAS.329..897H} Hurley, J.~R., Tout, C.~A., \& Pols, O.~R.\ 2002, \mnras, 329, 4, 897. doi:10.1046/j.1365-8711.2002.05038.x
\bibitem[Hurley et al.(2000)]{2000MNRAS.315..543H} Hurley, J.~R., Pols, O.~R., \& Tout, C.~A.\ 2000, \mnras, 315, 3, 543. doi:10.1046/j.1365-8711.2000.03426.x
\bibitem[Hussain(2011)]{2011ASPC..447..143H} Hussain, G.~A.~J.\ 2011, Evolution of Compact Binaries, 447, 143. doi:10.48550/arXiv.1202.5075
\bibitem[Inight et al.(2023)]{2023MNRAS.525.3597I} Inight, K., G{\"a}nsicke, B.~T., Schwope, A., et al.\ 2023, \mnras, 525, 3, 3597. doi:10.1093/mnras/stad2409
\bibitem[Inight et al.(2023)]{2023MNRAS.524.4867I} Inight, K., G{\"a}nsicke, B.~T., Breedt, E., et al.\ 2023, \mnras, 524, 4, 4867. doi:10.1093/mnras/stad2018
\bibitem[Istrate et al.(2014)]{2014A&A...571A..45I} Istrate, A.~G., Tauris, T.~M., \& Langer, N.\ 2014, \aap, 571, A45. doi:10.1051/0004-6361/201424680
\bibitem[Jao et al.(2022)]{2022ApJ...940..145J} Jao, W.-C., Couperus, A.~A., Vrijmoet, E.~H., et al.\ 2022, \apj, 940, 2, 145. doi:10.3847/1538-4357/ac9cd8
\bibitem[K{\'a}ra et al.(2025)]{2025A&A...699A..81K} K{\'a}ra, J., Zharikov, S., Wolf, M., et al.\ 2025, \aap, 699, A81. doi:10.1051/0004-6361/202553970
\bibitem[King \& Kolb(1995)]{1995ApJ...439..330K} King, A.~R. \& Kolb, U.\ 1995, \apj, 439, 330. doi:10.1086/175176
\bibitem[Knigge et al.(2011)]{2011ApJS..194...28K} Knigge, C., Baraffe, I., \& Patterson, J.\ 2011, \apjs, 194, 2, 28. doi:10.1088/0067-0049/194/2/28
\bibitem[Kolb(1993)]{1993A&A...271..149K} Kolb, U.\ 1993, \aap, 271, 149. 
\bibitem[Kolb et al.(2001)]{2001ApJ...563..958K} Kolb, U., Rappaport, S., Schenker, K., et al.\ 2001, \apj, 563, 2, 958. doi:10.1086/324074
\bibitem[Kroupa et al.(1993)]{1993MNRAS.262..545K} Kroupa, P., Tout, C.~A., \& Gilmore, G.\ 1993, \mnras, 262, 545. doi:10.1093/mnras/262.3.545
\bibitem[Landau \& Lifshitz(1975)]{1975ctf..book.....L} Landau, L.~D. \& Lifshitz, E.~M.\ 1975, . 
\bibitem[Landin et al.(2023)]{2023MNRAS.519.5304L} Landin, N.~R., Mendes, L.~T.~S., Vaz, L.~P.~R., et al.\ 2023, \mnras, 519, 4, 5304. doi:10.1093/mnras/stac3823
\bibitem[Lee et al.(2024)]{2024ApJ...964..186L} Lee, Y., Moon, D.-S., Kim, S.~C., et al.\ 2024, \apj, 964, 2, 186. doi:10.3847/1538-4357/ad25ff
\bibitem[Ma et al.(2013)]{2013ApJ...778L..32M} Ma, X., Chen, X., Chen, H.-. liang ., et al.\ 2013, \apjl, 778, L32. doi:10.1088/2041-8205/778/2/L32
\bibitem[Macrie et al.(2024)]{2024RNAAS...8..299M} Macrie, C.~W., Rivera Sandoval, L., Cavecchi, Y., et al.\ 2024, Research Notes of the American Astronomical Society, 8, 12, 299. doi:10.3847/2515-5172/ad9a64
\bibitem[Magaudda et al.(2020)]{2020A&A...638A..20M} Magaudda, E., Stelzer, B., Covey, K.~R., et al.\ 2020, \aap, 638, A20. doi:10.1051/0004-6361/201937408
\bibitem[Matt et al.(2012)]{2012ApJ...754L..26M} Matt, S.~P., MacGregor, K.~B., Pinsonneault, M.~H., et al.\ 2012, \apjl, 754, 2, L26. doi:10.1088/2041-8205/754/2/L26
\bibitem[Matt et al.(2015)]{2015ApJ...799L..23M} Matt, S.~P., Brun, A.~S., Baraffe, I., et al.\ 2015, \apjl, 799, 2, L23. doi:10.1088/2041-8205/799/2/L23
\bibitem[McAllister et al.(2019)]{2019MNRAS.486.5535M} McAllister, M., Littlefair, S.~P., Parsons, S.~G., et al.\ 2019, \mnras, 486, 4, 5535. doi:10.1093/mnras/stz976
\bibitem[Mestel(1968)]{1968MNRAS.138..359M} Mestel, L.\ 1968, \mnras, 138, 359. doi:10.1093/mnras/138.3.359
\bibitem[Medina et al.(2020)]{2020ApJ...905..107M} Medina, A.~A., Winters, J.~G., Irwin, J.~M., et al.\ 2020, \apj, 905, 2, 107. doi:10.3847/1538-4357/abc686
\bibitem[Pala et al.(2017)]{2017MNRAS.466.2855P} Pala, A.~F., G{\"a}nsicke, B.~T., Townsley, D., et al.\ 2017, \mnras, 466, 3, 2855. doi:10.1093/mnras/stw3293
\bibitem[Pala et al.(2020)]{2020MNRAS.494.3799P} Pala, A.~F., G{\"a}nsicke, B.~T., Breedt, E., et al.\ 2020, \mnras, 494, 3, 3799. doi:10.1093/mnras/staa764
\bibitem[Pala et al.(2022)]{2022MNRAS.510.6110P} Pala, A.~F., G{\"a}nsicke, B.~T., Belloni, D., et al.\ 2022, \mnras, 510, 4, 6110. doi:10.1093/mnras/stab3449
\bibitem[Patterson(1998)]{1998PASP..110.1132P} Patterson, J.\ 1998, \pasp, 110, 752, 1132. doi:10.1086/316233
\bibitem[Paxton et al.(2019)]{2019ApJS..243...10P} Paxton, B., Smolec, R., Schwab, J., et al.\ 2019, \apjs, 243, 10. doi:10.3847/1538-4365/ab2241
\bibitem[Paxton et al.(2018)]{2018ApJS..234...34P} Paxton, B., Schwab, J., Bauer, E.~B., et al.\ 2018, \apjs, 234, 34. doi:10.3847/1538-4365/aaa5a8
\bibitem[Paxton et al.(2015)]{2015ApJS..220...15P} Paxton, B., Marchant, P., Schwab, J., et al.\ 2015, \apjs, 220, 15. doi:10.1088/0067-0049/220/1/15
\bibitem[Paxton et al.(2013)]{2013ApJS..208....4P} Paxton, B., Cantiello, M., Arras, P., et al.\ 2013, \apjs, 208, 4. doi:10.1088/0067-0049/208/1/4
\bibitem[Paxton et al.(2011)]{2011ApJS..192....3P} Paxton, B., Bildsten, L., Dotter, A., et al.\ 2011, \apjs, 192, 3. doi:10.1088/0067-0049/192/1/3
\bibitem[Pfahl et al.(2003)]{2003ApJ...597.1036P} Pfahl, E., Rappaport, S., \& Podsiadlowski, P.\ 2003, \apj, 597, 2, 1036. doi:10.1086/378632
\bibitem[Podsiadlowski et al.(2002)]{2002ApJ...565.1107P} Podsiadlowski, P., Rappaport, S., \& Pfahl, E.~D.\ 2002, \apj, 565, 2, 1107. doi:10.1086/324686
\bibitem[Ramsay et al.(2018)]{2018A&A...620A.141R} Ramsay, G., Green, M.~J., Marsh, T.~R., et al.\ 2018, \aap, 620, A141. doi:10.1051/0004-6361/201834261
\bibitem[Rappaport et al.(1983)]{1983ApJ...275..713R} Rappaport, S., Verbunt, F., \& Joss, P.~C.\ 1983, \apj, 275, 713. doi:10.1086/161569
\bibitem[Reimers(1975)]{1975MSRSL...8..369R} Reimers, D.\ 1975, Memoires of the Societe Royale des Sciences de Liege, 8, 369. 
\bibitem[Reiners et al.(2009)]{2009ApJ...692..538R} Reiners, A., Basri, G., \& Browning, M.\ 2009, \apj, 692, 1, 538. doi:10.1088/0004-637X/692/1/538
\bibitem[Reiners \& Mohanty(2012)]{2012ApJ...746...43R} Reiners, A. \& Mohanty, S.\ 2012, \apj, 746, 1, 43. doi:10.1088/0004-637X/746/1/43
\bibitem[Reiners et al.(2022)]{2022A&A...662A..41R} Reiners, A., Shulyak, D., K{\"a}pyl{\"a}, P.~J., et al.\ 2022, \aap, 662, A41. doi:10.1051/0004-6361/202243251
\bibitem[Sadeghi Ardestani et al.(2017)]{2017MNRAS.472.2590S} Sadeghi Ardestani, L., Guillot, T., \& Morel, P.\ 2017, \mnras, 472, 3, 2590. doi:10.1093/mnras/stx2039
\bibitem[Sarkar et al.(2024)]{2024A&A...686L..19S} Sarkar, A., Rodriguez, A.~C., Ginzburg, S., et al.\ 2024, \aap, 686, L19. doi:10.1051/0004-6361/202449345
\bibitem[Scherbak \& Fuller(2023)]{2023MNRAS.518.3966S} Scherbak, P. \& Fuller, J.\ 2023, \mnras, 518, 3, 3966. doi:10.1093/mnras/stac3313
\bibitem[Schreiber et al.(2016)]{2016MNRAS.455L..16S} Schreiber, M.~R., Zorotovic, M., \& Wijnen, T.~P.~G.\ 2016, \mnras, 455, 1, L16. doi:10.1093/mnrasl/slv144
\bibitem[Schreiber et al.(2023)]{2023A&A...679L...8S} Schreiber, M.~R., Belloni, D., \& van Roestel, J.\ 2023, \aap, 679, L8. doi:10.1051/0004-6361/202347766
\bibitem[Sion(1995)]{1995ApJ...438..876S} Sion, E.~M.\ 1995, \apj, 438, 876. doi:10.1086/175129
\bibitem[Skumanich(1972)]{1972ApJ...171..565S} Skumanich, A.\ 1972, \apj, 171, 565. doi:10.1086/151310
\bibitem[Soethe \& Kepler(2021)]{2021MNRAS.506.3266S} Soethe, L.~T.~T. \& Kepler, S.~O.\ 2021, \mnras, 506, 3, 3266. doi:10.1093/mnras/stab1916
\bibitem[Shao \& Li(2014)]{2014ApJ...796...37S} Shao, Y. \& Li, X.-D.\ 2014, \apj, 796, 1, 37. doi:10.1088/0004-637X/796/1/37
\bibitem[Shao \& Li(2015)]{2015ApJ...809...99S} Shao, Y. \& Li, X.-D.\ 2015, \apj, 809, 1, 99. doi:10.1088/0004-637X/809/1/99
\bibitem[Shen \& Quataert(2022)]{2022ApJ...938...31S} Shen, K.~J. \& Quataert, E.\ 2022, \apj, 938, 1, 31. doi:10.3847/1538-4357/ac9136
\bibitem[Southworth et al.(2009)]{2009A&A...507..929S} Southworth, J., Hickman, R.~D.~G., Marsh, T.~R., et al.\ 2009, \aap, 507, 2, 929. doi:10.1051/0004-6361/200912885
\bibitem[Tang et al.(2024)]{2024ApJ...977...34T} Tang, W.-S., Li, X.-D., \& Cui, Z.\ 2024, \apj, 977, 1, 34. doi:10.3847/1538-4357/ad8880
\bibitem[Tang et al.(2025)]{2025ApJ...990..141T} Tang, W., Li, X.-D., \& Wang, B.\ 2025, \apj, 990, 2, 141. doi:10.3847/1538-4357/adf634
\bibitem[Tovmassian et al.(2025)]{2025arXiv250821358T} Tovmassian, G., Belloni, D., Pala, A.~F., et al.\ 2025, , arXiv:2508.21358. doi:10.48550/arXiv.2508.21358
\bibitem[Townsley \& Bildsten(2003)]{2003ApJ...596L.227T} Townsley, D.~M. \& Bildsten, L.\ 2003, \apjl, 596, 2, L227. doi:10.1086/379535
\bibitem[Townsley \& Bildsten(2004)]{2004ApJ...600..390T} Townsley, D.~M. \& Bildsten, L.\ 2004, \apj, 600, 1, 390. doi:10.1086/379647
\bibitem[Townsley \& G{\"a}nsicke(2009)]{2009ApJ...693.1007T} Townsley, D.~M. \& G{\"a}nsicke, B.~T.\ 2009, \apj, 693, 1, 1007. doi:10.1088/0004-637X/693/1/1007
%\bibitem[Tovmassian et al.(2025)]{2025arXiv250821358T} Tovmassian, G., Belloni, D., Pala, A.~F., et al.\ 2025, , arXiv:2508.21358. doi:10.48550/arXiv.2508.21358
\bibitem[Van et al.(2019)]{2019MNRAS.483.5595V} Van, K.~X., Ivanova, N., \& Heinke, C.~O.\ 2019, \mnras, 483, 4, 5595. doi:10.1093/mnras/sty3489
%\bibitem[Van et al.(2019)]{2019MNRAS.483.5595V} Van, K.~X., Ivanova, N., \& Heinke, C.~O.\ 2019, \mnras, 483, 4, 5595. doi:10.1093/mnras/sty3489
\bibitem[Van \& Ivanova(2019)]{2019ApJ...886L..31V} Van, K.~X. \& Ivanova, N.\ 2019, \apjl, 886, 2, L31. doi:10.3847/2041-8213/ab571c
\bibitem[Verbunt \& Zwaan(1981)]{1981A&A...100L...7V} Verbunt, F. \& Zwaan, C.\ 1981, \aap, 100, L7. 
\bibitem[Wang et al.(2010)]{2010MNRAS.401.2729W} Wang, B., Li, X.-D., \& Han, Z.-W.\ 2010, \mnras, 401, 4, 2729. doi:10.1111/j.1365-2966.2009.15857.x
\bibitem[Wei et al.(2023)]{2023A&A...679A..74W} Wei, N., Jiang, L., \& Chen, W.-C.\ 2023, \aap, 679, A74. doi:10.1051/0004-6361/202346397
\bibitem[Wright et al.(2011)]{2011ApJ...743...48W} Wright, N.~J., Drake, J.~J., Mamajek, E.~E., et al.\ 2011, \apj, 743, 1, 48. doi:10.1088/0004-637X/743/1/48
\bibitem[Wright \& Drake(2016)]{2016Natur.535..526W} Wright, N.~J. \& Drake, J.~J.\ 2016, \nat, 535, 7613, 526. doi:10.1038/nature18638
\bibitem[Wu et al.(2017)]{2017A&A...604A..31W} Wu, C., Wang, B., Liu, D., et al.\ 2017, \aap, 604, A31. doi:10.1051/0004-6361/201630099
\bibitem[Xu \& Li(2010)]{2010ApJ...716..114X} Xu, X.-J. \& Li, X.-D.\ 2010, \apj, 716, 1, 114. doi:10.1088/0004-637X/716/1/114
\bibitem[Xu \& Li(2018)]{2018MNRAS.480.3856X} Xu, X.-T. \& Li, X.-D.\ 2018, \mnras, 480, 3, 3856. doi:10.1093/mnras/sty2146
\bibitem[Yang \& Li(2024)]{2024ApJ...974..298Y} Yang, H.-R. \& Li, X.-D.\ 2024, \apj, 974, 2, 298. doi:10.3847/1538-4357/ad7824
\bibitem[Zhou et al.(2026)]{2026A&A...707A..76Z} Zhou, B., Zhu, C., L{\"u}, G., et al.\ 2026, \aap, 707, A76. doi:10.1051/0004-6361/202556958
\bibitem[Zorotovic et al.(2011)]{2011A&A...536A..42Z} Zorotovic, M., Schreiber, M.~R., \& G{\"a}nsicke, B.~T.\ 2011, \aap, 536, A42. doi:10.1051/0004-6361/201116626
\bibitem[Zorotovic \& Schreiber(2022)]{2022MNRAS.513.3587Z} Zorotovic, M. \& Schreiber, M.\ 2022, \mnras, 513, 3, 3587. doi:10.1093/mnras/stac1137





%\bibitem[Mestel(1968)]{1968MNRAS.138..359M} Mestel, L.\ 1968, \mnras, 138, 359. doi:10.1093/mnras/138.3.359
%\bibitem[Verbunt \& Zwaan(1981)]{1981A&A...100L...7V} Verbunt, F. \& Zwaan, C.\ 1981, \aap, 100, L7. 
%\bibitem[Rappaport et al.(1983)]{1983ApJ...275..713R} Rappaport, S., Verbunt, F., \& Joss, P.~C.\ 1983, \apj, 275, 713. doi:10.1086/161569
%\bibitem[Podsiadlowski et al.(2002)]{2002ApJ...565.1107P} Podsiadlowski, P., Rappaport, S., \& Pfahl, E.~D.\ 2002, \apj, 565, 2, 1107. doi:10.1086/324686
%\bibitem[Pfahl et al.(2003)]{2003ApJ...597.1036P} Pfahl, E., Rappaport, S., \& Podsiadlowski, P.\ 2003, \apj, 597, 2, 1036. doi:10.1086/378632
%\bibitem[Shao \& Li(2015)]{2015ApJ...809...99S} Shao, Y. \& Li, X.-D.\ 2015, \apj, 809, 1, 99. doi:10.1088/0004-637X/809/1/99
%\bibitem[Van et al.(2019)]{2019MNRAS.483.5595V} Van, K.~X., Ivanova, N., \& Heinke, C.~O.\ 2019, \mnras, 483, 4, 5595. doi:10.1093/mnras/sty3489
%\bibitem[Deng et al.(2021)]{2021ApJ...909..174D} Deng, Z.-L., Li, X.-D., Gao, Z.-F., et al.\ 2021, \apj, 909, 2, 174. doi:10.3847/1538-4357/abe0b2
%\bibitem[Istrate et al.(2014)]{2014A&A...571A..45I} Istrate, A.~G., Tauris, T.~M., \& Langer, N.\ 2014, \aap, 571, A45. doi:10.1051/0004-6361/201424680
%\bibitem[Knigge et al.(2011)]{2011ApJS..194...28K} Knigge, C., Baraffe, I., \& Patterson, J.\ 2011, \apjs, 194, 2, 28. doi:10.1088/0067-0049/194/2/28
%\bibitem[Tang et al.(2025)]{2025ApJ...990..141T} Tang, W., Li, X.-D., \& Wang, B.\ 2025, \apj, 990, 2, 141. doi:10.3847/1538-4357/adf634
%\bibitem[Matt et al.(2012)]{2012ApJ...754L..26M} Matt, S.~P., MacGregor, K.~B., Pinsonneault, M.~H., et al.\ 2012, \apjl, 754, 2, L26. doi:10.1088/2041-8205/754/2/L26
%\bibitem[Reiners \& Mohanty(2012)]{2012ApJ...746...43R} Reiners, A. \& Mohanty, S.\ 2012, \apj, 746, 1, 43. doi:10.1088/0004-637X/746/1/43
%\bibitem[Van \& Ivanova(2019)]{2019ApJ...886L..31V} Van, K.~X. \& Ivanova, N.\ 2019, \apjl, 886, 2, L31. doi:10.3847/2041-8213/ab571c
%\bibitem[Van et al.(2019)]{2019MNRAS.483.5595V} Van, K.~X., Ivanova, N., \& Heinke, C.~O.\ 2019, \mnras, 483, 4, 5595. doi:10.1093/mnras/sty3489
%\bibitem[Echeveste et al.(2024)]{2024MNRAS.530.4277E} Echeveste, M., Novarino, M.~L., Benvenuto, O.~G., et al.\ 2024, \mnras, 530, 4, 4277. doi:10.1093/mnras/stae1115
%\bibitem[Yang \& Li(2024)]{2024ApJ...974..298Y} Yang, H.-R. \& Li, X.-D.\ 2024, \apj, 974, 2, 298. doi:10.3847/1538-4357/ad7824
%\bibitem[Chen et al.(2021)]{2021MNRAS.503.3540C} Chen, H.-L., Tauris, T.~M., Han, Z., et al.\ 2021, \mnras, 503, 3, 3540. doi:10.1093/mnras/stab670
%\bibitem[Tovmassian et al.(2025)]{2025arXiv250821358T} Tovmassian, G., Belloni, D., Pala, A.~F., et al.\ 2025, , arXiv:2508.21358. doi:10.48550/arXiv.2508.21358
%\bibitem[Soethe \& Kepler(2021)]{2021MNRAS.506.3266S} Soethe, L.~T.~T. \& Kepler, S.~O.\ 2021, \mnras, 506, 3, 3266. doi:10.1093/mnras/stab1916
%\bibitem[Belloni \& Schreiber(2023)]{2023A&A...678A..34B} Belloni, D. \& Schreiber, M.~R.\ 2023, \aap, 678, A34. doi:10.1051/0004-6361/202347047
%\bibitem[Belloni et al.(2025)]{2025A&A...697A.100B} Belloni, D., Schreiber, M.~R., \& El-Badry, K.\ 2025, \aap, 697, A100. doi:10.1051/0004-6361/202553937
%\bibitem[Fan et al.(2024)]{2024ApJ...976..210F} Fan, Y.-N., Shao, Y., \& Chen, W.-C.\ 2024, \apj, 976, 2, 210. doi:10.3847/1538-4357/ad8b48
%\bibitem[Wei et al.(2023)]{2023A&A...679A..74W} Wei, N., Jiang, L., \& Chen, W.-C.\ 2023, \aap, 679, A74. doi:10.1051/0004-6361/202346397
%\bibitem[Deng \& Li(2024)]{2024ApJ...971...54D} Deng, Z.-L. \& Li, X.-D.\ 2024, \apj, 971, 1, 54. doi:10.3847/1538-4357/ad5fec
%\bibitem[Barraza-Jorquera et al.(2025)]{2025A&A...696A..92B} Barraza-Jorquera, J.~A., Schreiber, M.~R., \& Belloni, D.\ 2025, \aap, 696, A92. doi:10.1051/0004-6361/202553757
%\bibitem[Belloni et al.(2024)]{2024A&A...682A..33B} Belloni, D., Schreiber, M.~R., Moe, M., et al.\ 2024, \aap, 682, A33. doi:10.1051/0004-6361/202347931
%\bibitem[Pala et al.(2022)]{2022MNRAS.510.6110P} Pala, A.~F., G{\"a}nsicke, B.~T., Belloni, D., et al.\ 2022, \mnras, 510, 4, 6110. doi:10.1093/mnras/stab3449
%\bibitem[Zorotovic et al.(2011)]{2011A&A...536A..42Z} Zorotovic, M., Schreiber, M.~R., \& G{\"a}nsicke, B.~T.\ 2011, \aap, 536, A42. doi:10.1051/0004-6361/201116626
%\bibitem[Paxton et al.(2019)]{2019ApJS..243...10P} Paxton, B., Smolec, R., Schwab, J., et al.\ 2019, \apjs, 243, 10. doi:10.3847/1538-4365/ab2241
%\bibitem[Paxton et al.(2018)]{2018ApJS..234...34P} Paxton, B., Schwab, J., Bauer, E.~B., et al.\ 2018, \apjs, 234, 34. doi:10.3847/1538-4365/aaa5a8
%\bibitem[Paxton et al.(2015)]{2015ApJS..220...15P} Paxton, B., Marchant, P., Schwab, J., et al.\ 2015, \apjs, 220, 15. doi:10.1088/0067-0049/220/1/15
%\bibitem[Paxton et al.(2013)]{2013ApJS..208....4P} Paxton, B., Cantiello, M., Arras, P., et al.\ 2013, \apjs, 208, 4. doi:10.1088/0067-0049/208/1/4
%\bibitem[Paxton et al.(2011)]{2011ApJS..192....3P} Paxton, B., Bildsten, L., Dotter, A., et al.\ 2011, \apjs, 192, 3. doi:10.1088/0067-0049/192/1/3
%\bibitem[Landau \& Lifshitz(1975)]{1975ctf..book.....L} Landau, L.~D. \& Lifshitz, E.~M.\ 1975, . 
%\bibitem[Kolb et al.(2001)]{2001ApJ...563..958K} Kolb, U., Rappaport, S., Schenker, K., et al.\ 2001, \apj, 563, 2, 958. doi:10.1086/324074
%\bibitem[Tang et al.(2024)]{2024ApJ...977...34T} Tang, W.-S., Li, X.-D., \& Cui, Z.\ 2024, \apj, 977, 1, 34. doi:10.3847/1538-4357/ad8880
%\bibitem[Shen \& Quataert(2022)]{2022ApJ...938...31S} Shen, K.~J. \& Quataert, E.\ 2022, \apj, 938, 1, 31. doi:10.3847/1538-4357/ac9136
%\bibitem[Wu et al.(2017)]{2017A&A...604A..31W} Wu, C., Wang, B., Liu, D., et al.\ 2017, \aap, 604, A31. doi:10.1051/0004-6361/201630099
%\bibitem[Wang et al.(2010)]{2010MNRAS.401.2729W} Wang, B., Li, X.-D., \& Han, Z.-W.\ 2010, \mnras, 401, 4, 2729. doi:10.1111/j.1365-2966.2009.15857.x
%\bibitem[Ma et al.(2013)]{2013ApJ...778L..32M} Ma, X., Chen, X., Chen, H.-. liang ., et al.\ 2013, \apjl, 778, L32. doi:10.1088/2041-8205/778/2/L32
%\bibitem[Skumanich(1972)]{1972ApJ...171..565S} Skumanich, A.\ 1972, \apj, 171, 565. doi:10.1086/151310
%\bibitem[Reimers(1975)]{1975MSRSL...8..369R} Reimers, D.\ 1975, Memoires of the Societe Royale des Sciences de Liege, 8, 369. 
%\bibitem[Andronov et al.(2003)]{2003ApJ...582..358A} Andronov, N., Pinsonneault, M., \& Sills, A.\ 2003, \apj, 582, 1, 358. doi:10.1086/343030
%\bibitem[Chaboyer et al.(1995)]{1995ApJ...441..865C} Chaboyer, B., Demarque, P., \& Pinsonneault, M.~H.\ 1995, \apj, 441, 865. doi:10.1086/175408
%\bibitem[Hussain(2011)]{2011ASPC..447..143H} Hussain, G.~A.~J.\ 2011, Evolution of Compact Binaries, 447, 143. doi:10.48550/arXiv.1202.5075
%\bibitem[Inight et al.(2023)]{2023MNRAS.525.3597I} Inight, K., G{\"a}nsicke, B.~T., Schwope, A., et al.\ 2023, \mnras, 525, 3, 3597. doi:10.1093/mnras/stad2409
%\bibitem[Patterson(1998)]{1998PASP..110.1132P} Patterson, J.\ 1998, \pasp, 110, 752, 1132. doi:10.1086/316233
%\bibitem[Pala et al.(2017)]{2017MNRAS.466.2855P} Pala, A.~F., G{\"a}nsicke, B.~T., Townsley, D., et al.\ 2017, \mnras, 466, 3, 2855. doi:10.1093/mnras/stw3293
%\bibitem[Kolb(1993)]{1993A&A...271..149K} Kolb, U.\ 1993, \aap, 271, 149. 
%\bibitem[Belloni et al.(2018)]{2018MNRAS.478.5626B} Belloni, D., Schreiber, M.~R., Zorotovic, M., et al.\ 2018, \mnras, 478, 4, 5626. doi:10.1093/mnras/sty1421
%\bibitem[McAllister et al.(2019)]{2019MNRAS.486.5535M} McAllister, M., Littlefair, S.~P., Parsons, S.~G., et al.\ 2019, \mnras, 486, 4, 5535. doi:10.1093/mnras/stz976
%\bibitem[Southworth et al.(2009)]{2009A&A...507..929S} Southworth, J., Hickman, R.~D.~G., Marsh, T.~R., et al.\ 2009, \aap, 507, 2, 929. doi:10.1051/0004-6361/200912885
%\bibitem[Inight et al.(2023)]{2023MNRAS.524.4867I} Inight, K., G{\"a}nsicke, B.~T., Breedt, E., et al.\ 2023, \mnras, 524, 4, 4867. doi:10.1093/mnras/stad2018
%\bibitem[Kroupa et al.(1993)]{1993MNRAS.262..545K} Kroupa, P., Tout, C.~A., \& Gilmore, G.\ 1993, \mnras, 262, 545. doi:10.1093/mnras/262.3.545
%\bibitem[Xu \& Li(2010)]{2010ApJ...716..114X} Xu, X.-J. \& Li, X.-D.\ 2010, \apj, 716, 1, 114. doi:10.1088/0004-637X/716/1/114
%\bibitem[Shao \& Li(2014)]{2014ApJ...796...37S} Shao, Y. \& Li, X.-D.\ 2014, \apj, 796, 1, 37. doi:10.1088/0004-637X/796/1/37
%\bibitem[Zorotovic \& Schreiber(2022)]{2022MNRAS.513.3587Z} Zorotovic, M. \& Schreiber, M.\ 2022, \mnras, 513, 3, 3587. doi:10.1093/mnras/stac1137
%\bibitem[Scherbak \& Fuller(2023)]{2023MNRAS.518.3966S} Scherbak, P. \& Fuller, J.\ 2023, \mnras, 518, 3, 3966. doi:10.1093/mnras/stac3313
%\bibitem[Pala et al.(2020)]{2020MNRAS.494.3799P} Pala, A.~F., G{\"a}nsicke, B.~T., Breedt, E., et al.\ 2020, \mnras, 494, 3, 3799. doi:10.1093/mnras/staa764
%\bibitem[Green et al.(2025)]{2025A&A...700A.107G} Green, M.~J., van Roestel, J., \& Wong, T.~L.~S.\ 2025, \aap, 700, A107. doi:10.1051/0004-6361/202554925
%\bibitem[K{\'a}ra et al.(2025)]{2025A&A...699A..81K} K{\'a}ra, J., Zharikov, S., Wolf, M., et al.\ 2025, \aap, 699, A81. doi:10.1051/0004-6361/202553970
%\bibitem[Lee et al.(2024)]{2024ApJ...964..186L} Lee, Y., Moon, D.-S., Kim, S.~C., et al.\ 2024, \apj, 964, 2, 186. doi:10.3847/1538-4357/ad25ff
%\bibitem[Wright \& Drake(2016)]{2016Natur.535..526W} Wright, N.~J. \& Drake, J.~J.\ 2016, \nat, 535, 7613, 526. doi:10.1038/nature18638
%\bibitem[Reiners et al.(2022)]{2022A&A...662A..41R} Reiners, A., Shulyak, D., K{\"a}pyl{\"a}, P.~J., et al.\ 2022, \aap, 662, A41. doi:10.1051/0004-6361/202243251
%\bibitem[Wright et al.(2011)]{2011ApJ...743...48W} Wright, N.~J., Drake, J.~J., Mamajek, E.~E., et al.\ 2011, \apj, 743, 1, 48. doi:10.1088/0004-637X/743/1/48
%\bibitem[Landin et al.(2023)]{2023MNRAS.519.5304L} Landin, N.~R., Mendes, L.~T.~S., Vaz, L.~P.~R., et al.\ 2023, \mnras, 519, 4, 5304. doi:10.1093/mnras/stac3823
%\bibitem[Gossage et al.(2025)]{2025ApJ...988..102G} Gossage, S., Kiman, R., Monsch, K., et al.\ 2025, \apj, 988, 1, 102. doi:10.3847/1538-4357/adde4d
%\bibitem[Jao et al.(2022)]{2022ApJ...940..145J} Jao, W.-C., Couperus, A.~A., Vrijmoet, E.~H., et al.\ 2022, \apj, 940, 2, 145. doi:10.3847/1538-4357/ac9cd8
%\bibitem[Sarkar et al.(2024)]{2024A&A...686L..19S} Sarkar, A., Rodriguez, A.~C., Ginzburg, S., et al.\ 2024, \aap, 686, L19. doi:10.1051/0004-6361/202449345
%\bibitem[Schreiber et al.(2023)]{2023A&A...679L...8S} Schreiber, M.~R., Belloni, D., \& van Roestel, J.\ 2023, \aap, 679, L8. doi:10.1051/0004-6361/202347766
%\bibitem[Deng \& Li(2024)]{2024ApJ...971...54D} Deng, Z.-L. \& Li, X.-D.\ 2024, \apj, 971, 1, 54. doi:10.3847/1538-4357/ad5fec
%\bibitem[Tovmassian et al.(2025)]{2025arXiv250821358T} Tovmassian, G., Belloni, D., Pala, A.~F., et al.\ 2025, , arXiv:2508.21358. doi:10.48550/arXiv.2508.21358
%\bibitem[El-Badry et al.(2022)]{2022MNRAS.517.4916E} El-Badry, K., Conroy, C., Fuller, J., et al.\ 2022, \mnras, 517, 4, 4916. doi:10.1093/mnras/stac2945
%\bibitem[King \& Kolb(1995)]{1995ApJ...439..330K} King, A.~R. \& Kolb, U.\ 1995, \apj, 439, 330. doi:10.1086/175176
%\bibitem[Magaudda et al.(2020)]{2020A&A...638A..20M} Magaudda, E., Stelzer, B., Covey, K.~R., et al.\ 2020, \aap, 638, A20. doi:10.1051/0004-6361/201937408
%\bibitem[Reiners et al.(2009)]{2009ApJ...692..538R} Reiners, A., Basri, G., \& Browning, M.\ 2009, \apj, 692, 1, 538. doi:10.1088/0004-637X/692/1/538
%\bibitem[Medina et al.(2020)]{2020ApJ...905..107M} Medina, A.~A., Winters, J.~G., Irwin, J.~M., et al.\ 2020, \apj, 905, 2, 107. doi:10.3847/1538-4357/abc686
%\bibitem[Sion(1995)]{1995ApJ...438..876S} Sion, E.~M.\ 1995, \apj, 438, 876. doi:10.1086/175129
%\bibitem[Townsley \& Bildsten(2003)]{2003ApJ...596L.227T} Townsley, D.~M. \& Bildsten, L.\ 2003, \apjl, 596, 2, L227. doi:10.1086/379535
%\bibitem[Townsley \& Bildsten(2004)]{2004ApJ...600..390T} Townsley, D.~M. \& Bildsten, L.\ 2004, \apj, 600, 1, 390. doi:10.1086/379647
%\bibitem[Townsley \& G{\"a}nsicke(2009)]{2009ApJ...693.1007T} Townsley, D.~M. \& G{\"a}nsicke, B.~T.\ 2009, \apj, 693, 1, 1007. doi:10.1088/0004-637X/693/1/1007
%\bibitem[Hurley et al.(2002)]{2002MNRAS.329..897H} Hurley, J.~R., Tout, C.~A., \& Pols, O.~R.\ 2002, \mnras, 329, 4, 897. doi:10.1046/j.1365-8711.2002.05038.x
%\bibitem[Hurley et al.(2000)]{2000MNRAS.315..543H} Hurley, J.~R., Pols, O.~R., \& Tout, C.~A.\ 2000, \mnras, 315, 3, 543. doi:10.1046/j.1365-8711.2000.03426.x
%\bibitem[Xu \& Li(2018)]{2018MNRAS.480.3856X} Xu, X.-T. \& Li, X.-D.\ 2018, \mnras, 480, 3, 3856. doi:10.1093/mnras/sty2146
%\bibitem[Sadeghi Ardestani et al.(2017)]{2017MNRAS.472.2590S} Sadeghi Ardestani, L., Guillot, T., \& Morel, P.\ 2017, \mnras, 472, 3, 2590. doi:10.1093/mnras/stx2039

\end{thebibliography}
\end{document}